\documentclass[final,3p,times]{elsarticle}
\usepackage{amssymb}
\usepackage{color}

\journal{Journal of Computational Physics}

\begin{document}
\begin{frontmatter}

\title{Multidimensional Hall magnetohydrodynamics with isotropic or anisotropic
thermal pressure: numerical scheme and its validation using solitary waves}

\author[label1]{Marek Strumik\fnref{labelX}}
\ead{mstrumik@gmail.com}
\author[label2,label3]{Krzysztof Stasiewicz}

\address[label1]{Rudolf Peierls Centre for Theoretical Physics, University of Oxford, Oxford OX1 3NP, UK}
\address[label2]{Space Research Centre, Polish Academy of Sciences, Warsaw, Poland}
\address[label3]{Department of Physics and Astronomy, University of Zielona G\'ora, Zielona G\'ora, Poland}
\fntext[labelX]{on leave from Space Research Centre, Polish Academy of Sciences, Warsaw, Poland}

\begin{abstract}
We present a numerical solver for plasma dynamics simulations in Hall
magnetohydrodynamic (HMHD) approximation in one, two and three
dimensions. We consider both isotropic and anisotropic thermal
pressure cases, where a general gyrotropic approximation is used. Both
explicit energy conservation equation and general polytropic state
equations are considered. The numerical scheme incorporates second-order
Runge-Kutta advancing in time and Kurganov-Tadmor scheme with van Leer
flux limiter for the approximation of fluxes. A flux-interpolated
constrained-transport approach is used to preserve solenoidal magnetic
field in the simulations. The implemented code is validated using several
test problems previously described in the literature. Additionally, we
propose a new validation method for HMHD codes based on solitary waves
that provides a possibility of quantitative rigorous testing in nonlinear
(large amplitude) regime as an extension to standard tests using
small-amplitude whistler waves. Quantitative tests of accuracy and
performance of the implemented code show the fidelity of the proposed
approach.
\end{abstract}


\end{frontmatter}

\section{Introduction}

Hall magnetohydrodynamics (HMHD hereafter) provides a natural extension of
ideal or resistive magnetohydrodynamic (MHD) models for plasmas in the limit of small
scales, where the magnetic field is frozen into electron fluid but ions are
decoupled from the magnetic field lines \cite{Hub95}. As related to different masses
of ions and electrons, the inertial effects become important at scales of the order of
the ion inertial length (sometimes referred to as the ion skin depth)
$d_i=V_A/\Omega_{i}$, where $V_A$ is the Alfven speed and $\Omega_{i}$ is the
ion gyrofrequency. The HMHD physics is essentially contained in the Ohm's law
modified in comparison with the MHD formulation, which influences the transport
of the magnetic field in plasma through the Faraday's induction equation. The Hall term
also enters the energy conservation equation. Dispersive effects related to the Hall
term are responsible for the appearance of so-called whistler waves.
HMHD-related phenomena are studied as an important element of
fast magnetic reconnection \cite{Biretal01,MaBha01,HubRud04}. The HMHD
physics includes also processes of formation of solitary waves \cite{Sta04,Sta04b,Sta05,Stretal11}.
The Hall term is also important for modeling small-scale fluctuations in plasma
turbulence \cite{Minetal05,DmiMat06}.

In collisionless or weakly collisional plasmas one may expect the development
of thermal pressure anisotropies. Lack of collisional mechanisms of exchange
of particle energy between degrees of freedom parallel and
perpendicular to the magnetic field direction may obviously lead to an
asymmetric distribution function for particle velocities. In the
lowest-order approximation, a gyrotropic model of anisotropy applies,
where the distribution function is assumed to be bi-Maxwellian and
axially symmetric with respect to the local
magnetic field direction. In this approach, the parallel and perpendicular
temperatures are in general different and they evolve in time
in a different way. The pressure anisotropy is known to provide free energy for
the development of instabilities, that are believed to control the pressure
anisotropy in space plasmas as measured in-situ in the solar wind
\cite{Kasetal02,Heletal06,Matetal07,Baletal09}. Questions related to the
pressure anisotropy regulation in space plasmas have been investigated
extensively in various astrophysical aspects
\cite{Schetal05,HelTra08,Schetal08,Kunetal14,Seretal14}.

There exist a number of numerical codes for numerical simulations within
the HMHD framework. The codes use explicit time advancing (e.g. \cite{Hub03}) or implicit
scheme (e.g. \cite{ChaKno03,Lavetal09}). Efforts have been made towards
including adaptive mesh refinement in HMHD simulations \cite{Totetal08}. However,
quantitative validation of HMHD codes in nonlinear regime
is difficult due to the lack of analytic or semi-analytic problems that could be used
for this purpose. Quantitative testing of the accuracy of HMHD codes
consists mainly in studying of propagation of small-amplitude whistler
waves in the computational domain. To our knowledge, no general method of testing of
absolute accuracy has been proposed for the nonlinear regime of HMHD
dynamics.

In this paper, we discuss a method of solving of the HMHD equations
with the isotropic or anisotropic thermal pressure. The algorithm can be briefly
described as using the second-order Runge-Kutta advancing in time
and Kurganov-Tadmor scheme with van Leer flux limiter for the approximation of fluxes.
To preserve solenoidal
magnetic field during time evolution, the magnetic field transport equation
is advanced in time using so-called flux-interpolated constrained-transport
approach. The pressure tensor can be modeled in a gyrotropic approximation
with polytropic relations describing the evolution of the
parallel and perpendicular pressures. It is also possible to use
an equation for the evolution of the perpendicular pressure and
the explicit energy conservation equation, which guarantees the conservation of the
total energy averaged over the simulation box to a very high accuracy.
For isotropic pressure case also a polytropic state equation or
the explicit energy conservation equation can be used.
The presented scheme is intended for simulations of phenomena in the range of
scales of the order of the ion inertial length and larger. This range
of scales is determined by a general physical regime of validity of the HMHD equations,
but also by the explicit character of the proposed numerical scheme that
imposes strong constraints on the simulation time step. The algorithm is shown to
work properly for one-, two- and three-dimensional test problems of
different types: solitary waves propagation, magnetic reconnection, and the growth
of the firehose instability.
In this paper, we also discuss thoroughly a new testing method based on the propagation of solitary
structures as a possible testing framework for HMHD in the nonlinear regime.

\section{Physical model}
\subsection{HMHD equations in conservative form}
The following equations can be derived as describing plasma dynamics on scales comparable to
the ion inertial length scale in the collisionless plasma regime within
fluid approximation (see, e.g. Refs. \cite{Hub95,Sta05,KraTri73} for details). The mass and
momentum transport can be calculated by the following equations
\begin{equation}
\frac{\partial \rho}{\partial t} = - \nabla \cdot (\rho \mathbf{u})
\label{eq:dndt}
\end{equation}
and
\begin{equation}
\rho \left[ \frac{\partial \mathbf{u}}{\partial t} +
(\mathbf{u} \cdot \nabla) \mathbf{u} \right] =
\mathbf{J} \times \mathbf{B} - \nabla \cdot \mathbb{P},
\label{eq:momentum}
\end{equation}
correspondingly. The Ampere's law defines the current density $\mathbf{J}
= \mu_0^{-1} \nabla \times \mathbf{B}$, $\mathbb{P}_{ij}=p_{\perp} \delta_{ij} +
(p_{\parallel}-p_{\perp})B_i B_j/B^2$ is the pressure tensor (gyrotropic
approximation, $\parallel$ and $\bot$ directions are defined with respect
to the local magnetic field direction), $\rho=N m_{\mathrm{i}}$ is the
proton density, $N$ is the proton number density, $\mathbf{u}$ is the plasma velocity vector, $\mathbf{B}$ is the
magnetic field vector, $m_{\mathrm{i}}$ is the proton mass. The generalized Ohm's equation
\begin{equation}
-\mathbf{u}_\mathrm{H} \times \mathbf{B} = \mathbf{E} + \mathbf{u} \times \mathbf{B} - \eta \mathbf{J}
\label{eq:ohm}
\end{equation}
contains a Hall term on the left-hand side, where $\mathbf{u}_\mathrm{H}=
-\mathbf{J}/e N$ is a Hall velocity vector, $e$ is the proton
charge. The resistive term $\eta \mathbf{J}$ allows to incorporate 
effects of finite resistivity in the model, where $\eta$ formally denotes the magnetic diffusivity.
The above equations can be obtained formally from the kinetic Vlasov equation using a
standard procedure based on subsequent moments of the velocity
distribution function, where all terms proportional to the electron
inertial length are neglected \cite{KraTri73}. Additionally, we assumed
here a small electron temperature since otherwise an additional term
proportional to the gradient of the electron pressure $\nabla p_e/e N$
would have been required in Eq. (\ref{eq:ohm}). The electron pressure term could be
incorporated into the model in a simplified way (scalar pressure evolution by using isothermal or polytropic equation of state),
but a more elaborated approach with anisotropy of the electron pressure is presumably advantageous at least for some
problems, like e.g. magnetic reconnection process as recently suggested \cite{Leetal09,Egeetal13}. In our discussion,
the electron pressure term is neglected for simplicity (which corresponds to cold electrons limit), and its implementation
is deferred to future work.
Using Eq. (\ref{eq:ohm}), the Faraday's law
\begin{equation}
\frac{\partial \mathbf{B}}{\partial t} = - \nabla \times \mathbf{E}
\label{eq:faraday}
\end{equation}
and $\nabla \cdot \mathbf{B}=0$ condition
we can derive a transport equation for the magnetic field vector in
the conservative form
\begin{equation}
\frac{\partial \mathbf{B}}{\partial t} = - \nabla \cdot \left[
(\mathbf{u}+\mathbf{u}_\mathrm{H}) \mathbf{B} -
\mathbf{B} (\mathbf{u}+\mathbf{u}_\mathrm{H}) \right]
+ \eta \nabla^2 \mathbf{B}.
\label{eq:dbdt}
\end{equation}
Eq. (\ref{eq:momentum}) can be also rewritten in the conservative form
\begin{equation}
\frac{\partial (\rho \mathbf{\mathbf{u}})}{\partial t} =
- \nabla \cdot \left( \rho \mathbf{u}\mathbf{u} + \mathbb{P} -
\frac{\mathbf{B}\mathbf{B}}{\mu_0} +
\frac{B^2}{2 \mu_0}\mathbb{I} \right),
\label{eq:dvdt}
\end{equation}
where $\mathbb{I}=\delta_{ij}$ is the identity matrix. The energy conservation equation
reads as follows
\begin{equation}
\hspace{-0.5cm}
\frac{\partial \widetilde{E}}{\partial t} = - \nabla \cdot \left[ \left( \widetilde{E} + 
 \frac{B^2}{2 \mu_0} \right) \mathbf{u}
- ( \mathbf{u}_\mathrm{H} \cdot \mathbf{B} + \mathbf{u} \cdot \mathbf{B} )
\mathbf{B} + \frac{B^2}{\mu_0} \mathbf{u}_\mathrm{H} + \mathbb{P} \cdot \mathbf{u}
+ \eta \mathbf{J} \times \mathbf{B}
\right],
\label{eq:dedt}
\end{equation}
where the total energy density is $\widetilde{E}=\rho u^2/2 +
p/(\gamma-1) + B^2/2\mu_0$. We have assumed here that the thermal energy
density $\epsilon=p/(\gamma-1)$ ($\gamma=5/3$ is used hereafter) can be
defined by a scalar pressure $p=(2 p_\bot + p_\parallel)/3$
(one-third of the trace of the pressure tensor $\mathbb{P}$) \cite{Bit04}.

The formulation described above allows to study the effects of the
anisotropic pressure in HMHD. Since Eq. (\ref{eq:dedt}) in general does
not give the time evolution of $p_\bot$ and $p_\parallel$ separately, it
is necessary to make further assumptions regarding, e.g. a constant
pressure anisotropy or polytropic form of state equations for the perpendicular
and/or the parallel components of the pressure tensor. This leads to
conservation of the following quantity
\begin{equation}
\widetilde{S}=\frac{\widetilde{p}}{\rho^{\widetilde{\gamma}} B^{\widetilde{\kappa}}}
\end{equation}
in the plasma frame (along pathlines) for each pressure component $\widetilde{p}$, i.e.
\begin{equation}
\frac{\mathrm{d} \widetilde{S}}{\mathrm{d} t}=\frac{\partial \widetilde{S}}{\partial t}+(\mathbf{u} \cdot \nabla)\widetilde{S}=0.
\label{eq:dsdtp}
\end{equation}
Using the continuity equation (\ref{eq:dndt}) the above condition can be rewritten in the conservative form
\begin{equation}
\frac{\partial S}{\partial t} = - \nabla \cdot (S \mathbf{u}),
\label{eq:dsdtc}
\end{equation}
where $S=\rho \widetilde{S}$. It is convenient to
assume, e.g. that
\begin{equation}
p_\bot \propto \rho B^{\gamma_\bot-1}, \quad p_\parallel \propto \rho^{\gamma_\parallel} B^{1-\gamma_\parallel}
\label{eq:phauetal93}
\end{equation}
as suggested in Ref. \cite{Hauetal93}. Neglecting heat fluxes,
viscous and Joule's heating, and assuming that the time evolutions of
the parallel and perpendicular pressures are decoupled
lead to a well-known double-adiabatic (or the CGL)
approximation, with $\gamma_\bot=2$,
$\gamma_\parallel=3$ \cite{Cheetal57}. Alternatively, for $\gamma_\bot=1$ and
$\gamma_\parallel=1$ a double isothermal behavior is obtained.
One should note that the double-adiabatic and double-isothermal models are two special
cases of an entire family of polytropic models described by Eq. (\ref{eq:phauetal93}).

Using polytropic state equations in the conservative form of Eq. (\ref{eq:dsdtc})
for both the parallel and perpendicular pressures and evolving them independently
may lead to conservation of the total energy density with limited accuracy in the
simulation. The limited accuracy can be insufficient for
some problems, where energy density components: kinetic $\rho u^2/2$, thermal $3 p/2$ and 
magnetic $B^2/2 \mu_0$ differ by several orders of magnitude. For these
problems, it is advantageous to use only one polytropic equation (for example for
the perpendicular pressure) in the form of Eq. (\ref{eq:dsdtc}) and the energy conservation equation in
the explicit form of Eq. (\ref{eq:dedt}) to compute the second pressure
component from the total energy density $\widetilde{E}$. This approach guarantees the conservation
of the total energy density integrated over a periodic simulation box to
very high accuracy $\sim 10^{-12}$ if the total energy density is of order unity,
even in the presence of numerical errors introduced by discretization.

The equations outlined above constitute a gyrotropic HMHD model (gyrotropic
refers to the assumption of the symmetry of the pressure tensor with respect to the
local magnetic field direction).
Analysis of Eqs. (\ref{eq:dndt}) and (\ref{eq:dbdt})-(\ref{eq:dedt})
shows that the Hall term affects the magnetic field and energy transport
in the system, but does not influence directly the mass and momentum
transport. The Hall term introduces the dispersion scale length
related to the decoupling of the ion motion from
the magnetic field lines transport, while the electrons (due to their
smaller mass) remain frozen into the magnetic field lines. Resulting
difference in the ion and electron average velocities leads to the appearance of the Hall
term in the generalized Ohm's law of Eq. (\ref{eq:ohm}). If we set
$\mathbf{u}_\mathrm{H}=(0,0,0)$ in the above equations, we obtain the classical
MHD equations, where both ions and electrons are assumed to be
frozen into the transported magnetic field lines.
The equations presented above constitute a one-fluid approximation
that can be expected to be valid for magnetized plasma for spatial scales larger than the
ion inertial length (and Larmor radius) and time scales larger than the gyroperiod.
The electron pressure gradients are neglected in the presented approach, which formally
corresponds to cold electrons limit.
Possible extensions of the equations in the context of numerical simulations,
like e.g., two-fluid \cite{Haketal06} (including relativistic
effects \cite{Baletal16,Ama16}) or multi-fluid \cite{Wanetal15}
models have been considered in the literature.

\subsection{Normalization of HMHD equations}
The magnetic field and the number density are normalized to their
background values $B_0$ and $N_0$, correspondingly. The velocity is
normalized to the Alfven speed $V_\mathrm{A0}=B_0/\sqrt{\mu_0 \rho_0}$,
and the pressure is normalized to $P_0=B_0^2/\mu_0$. The length unit is
the ion inertial length $d_\mathrm{i}$ and the time unit is the
inverse of the proton cyclotron frequency $\Omega_\mathrm{i}^{-1}$.
The magnetic diffusivity $\eta$ is measured in $V_\mathrm{A0} d_\mathrm{i}$ units.
This procedure leads to the following normalized set of equations
in the conservative form
\begin{equation}
\frac{\partial N}{\partial t} = - \nabla \cdot (N \mathbf{u}),
\label{eq:dndtN}
\end{equation}
\begin{equation}
\frac{\partial (N \mathbf{\mathbf{u}})}{\partial t} =
- \nabla \cdot \left( N \mathbf{u}\mathbf{u} + \mathbb{P} -
\mathbf{B}\mathbf{B} + \frac{B^2}{2}\mathbb{I} \right),
\label{eq:dvdtN}
\end{equation}
\begin{equation}
\frac{\partial \mathbf{B}}{\partial t} = - \nabla \cdot \left[
(\mathbf{u}+\mathbf{u}_\mathrm{H}) \mathbf{B} -
\mathbf{B} (\mathbf{u}+\mathbf{u}_\mathrm{H}) \right]
+ \eta \nabla^2 \mathbf{B},
\label{eq:dbdtN}
\end{equation}
where $\mathbf{u}_\mathrm{H}= - \nabla \times \mathbf{B}/N$.
One can use the normalized energy equation
\begin{equation}
\frac{\partial \widetilde{E}}{\partial t} = - \nabla \cdot \left[ \left( \widetilde{E} + 
\frac{B^2}{2} \right) \mathbf{u}
- ( \mathbf{u}_\mathrm{H} \cdot \mathbf{B} + \mathbf{u} \cdot \mathbf{B} )
\mathbf{B} + B^2 \mathbf{u}_\mathrm{H} + \mathbb{P} \cdot \mathbf{u}
+ \eta \mathbf{J} \times \mathbf{B}
\right]
\label{eq:dedtN}
\end{equation}
or Eq. (\ref{eq:dsdtc}) for polytropic pressure relationships.
The normalized total energy density is
$\widetilde{E}=N u^2/2 + p /(\gamma-1) + B^2/2$. 

\subsection{Nonlinear wave solutions of the HMHD equations}
\label{sec:solit}

Classical methods of analysis of the HMHD equations in linear
regime lead to a dispersion relation describing
properties of small-amplitude wave solutions for the system
\cite{WanHau10}. An alternative approach \cite{Sta04,Sta04b,Sta05} makes
it possible to obtain exact nonlinear (large-amplitude) waves in the form
of solitary solutions that can be used to validate numerical solvers used
for time-dependent HMHD simulations. 

The nonlinear waves can be obtained as exact stationary
($\partial/\partial t=0$) solutions of Eqs.
(\ref{eq:dndtN})-(\ref{eq:dbdtN}) in 1.5-dimensional geometry (planar waves
approximation for three-dimensional fields, consistent with the assumption
$\partial/\partial y=\partial/\partial z=0$)
\cite{Sta04,Sta04b,Sta05}. In this approach we consider HMHD equations
in the wave frame, assuming that background (unperturbed) plasma moves 
with velocity $\mathbf{u}_0=(u_{x0}, 0, 0)$, the normalized background
density is $N_0=1$, and the background magnetic field vector is 
$\mathbf{B}_0=(\cos \! \alpha, 0, \sin \! \alpha)$, where $\alpha$ is the angle
between the magnetic field vector and the wave propagation direction $x$.
Using the above assumptions the condition $\nabla \cdot \mathbf{B}=0$
gives $B_x=\mathrm{const}=\cos \! \alpha$ and Eq. (\ref{eq:dndtN}) leads to
$N u_x=\mathrm{const}=u_{x0}$. The $x$-component of Eq. (\ref{eq:dvdtN})
(momentum conservation) can be then simplified to the following algebraic
equation for the isotropic pressure
\begin{equation}
\frac{u_{x0}^2}{N}+\frac{\beta}{2} N^\gamma + \frac{B^2}{2}=u_{x0}^2+\frac{\beta}{2} + \frac{1}{2},
\label{eq:nb}
\end{equation}
where $\beta=2 p_0/B_0^2$. A corresponding equation for the anisotropic pressure 
described by Eq. (\ref{eq:phauetal93}) reads as follows
\begin{equation}
\hspace{-0.5cm}
\frac{u_{x0}^2}{N}+\frac{\beta_\parallel}{2} N^{\gamma_\parallel} B^{-\gamma_\parallel-1} \cos^2 \! \alpha +
\frac{\beta_\bot}{2} N B^{\gamma_\bot-1} \left( 1-\frac{\cos^2 \! \alpha}{B^2} \right) +
\frac{B^2}{2}= u_{x0}^2+\frac{\beta_\parallel}{2} \cos^2 \! \alpha + \frac{\beta_\bot}{2} \left( 1-\cos^2 \! \alpha \right) + \frac{1}{2}.
\label{eq:nb_anis}
\end{equation}
Parameters $\beta_\parallel$ and $\beta_\bot$ are introduced to include different
thermal energy densities in the parallel and the perpendicular direction to the
magnetic field. The transverse ($y$ and $z$) components of Eq. (\ref{eq:dvdtN}) 
in stationary case can be expressed as
\begin{equation}
u_y = \frac{A B_y \cos \! \alpha}{u_{x0}}, \quad u_z = \frac{\cos \! \alpha}{u_{x0}} \left( A B_z - A_0 \sin \! \alpha \right).
\label{eq:vyvz}
\end{equation}
For the isotropic pressure $A=A_0=1$ and for the anisotropic pressure
$A=\left( 1 - \beta_\parallel N^{\gamma_\parallel} B^{-\gamma_\parallel-1}/2 + \beta_\bot N B^{\gamma_\bot-3}/2 \right)$ and $A_0=(1 - \beta_\parallel/2 + \beta_\bot/2 )$.
The transverse components of Eq. (\ref{eq:dbdtN}) (magnetic field transport) lead to
the ordinary differential equations (ODE)
\begin{equation}
\frac{\mathrm{d} B_y}{\mathrm{d} x} = - \frac{B_z u_{x0}}{\cos \!\alpha} +N \left( u_z + \frac{u_{x0} \sin \! \alpha}{\cos \! \alpha} \right), \quad
\frac{\mathrm{d} B_z}{\mathrm{d} x} = \frac{B_y u_{x0}}{\cos \! \alpha}-N u_y. 
\label{eq:dbidx}
\end{equation}
If we use Eq. (\ref{eq:vyvz}) to eliminate $u_y$ and $u_z$ from Eqs.
(\ref{eq:dbidx}), then on the right-hand side of Eqs. (\ref{eq:dbidx}) we
have functions dependent only on $B_y, B_z$ and $N$. The algebraic
equation (\ref{eq:nb}) (or Eq. (\ref{eq:nb_anis}) for the
anisotropic pressure) can be solved numerically to find $N=N(B)$,
$B=\sqrt{\cos^2 \! \alpha+B_y^2+B_z^2}$, thus we may conclude that the
right hand sides of Eqs. (\ref{eq:dbidx}) depend only on $B_y, B_z$.
Therefore we have a set of two coupled ODEs that can be solved
numerically to produce spatial profiles of waves that are stationary
solutions to the HMHD equations in the wave frame of reference.

\subsection{Linear analysis of wave solutions of the HMHD equations}

By decomposing variables into the background value and the fluctuation:
$N=N_0+n$, $B_y=B_{y0} + b_y$, $B_z= B_{z0} + b_z$ we can investigate
behavior of solutions of Eqs. (\ref{eq:dbidx}) in the proximity of the background
state $N_0=1, B_{y0}=0, B_{z0}=\sin \! \alpha$ that is a fixed point of
Eqs. (\ref{eq:dbidx}). Linear response of the density fluctuation $n$ to
the magnetic field perturbation ($b_y, b_z$) implied by Eqs.
(\ref{eq:nb}) or (\ref{eq:nb_anis}) is $n= b_z D \sin \! \alpha$, where
for the isotropic pressure
\begin{equation}
D=\frac{2}{2 u_{x0}^2-\beta \gamma}
\end{equation}
and for the anisotropic pressure
\begin{equation}
\hspace{-0.5cm}
D=\frac{-2+\beta_\bot+\cos^2 \! \alpha \, \left[ \beta_\parallel(\gamma_\parallel+1)+\beta_\bot(\gamma_\bot-3)\right]-\beta_\bot \gamma_\bot }
{-2 u_{x0}^2+\beta_\bot+\cos^2 \! \alpha \, (\beta_\parallel \gamma_\parallel-\beta_\bot)},
\end{equation}
correspondingly. Neglecting the terms of the second and higher order
(with respect to the fluctuations) we obtain the linearized system for
the isotropic pressure
\begin{equation}
\frac{\mathrm{d} b_y}{\mathrm{d} x} = \left( \frac{\cos \! \alpha}{u_{x0}} + \frac{(D \sin^2 \! \alpha \, - 1) u_{x0}}{\cos \! \alpha} \right) b_z, \quad
\frac{\mathrm{d} b_z}{\mathrm{d} x} = \left( \frac{u_{x0}}{\cos \! \alpha} - \frac{\cos \! \alpha}{u_{x0}} \right) b_y,
\label{eq:dbidx_lin_iso}
\end{equation}
and for the anisotropic pressure
\begin{eqnarray}
\label{eq:dbidx_lin_anis}
\frac{\mathrm{d} b_y}{\mathrm{d} x} & = & \bigg\{ \frac{(D \sin^2 \! \alpha -1) u_{x0}}{\cos \! \alpha} +
\frac{\cos \! \alpha \,\left[ 2+\beta_\bot-\beta_\parallel + \beta_\parallel \sin^2 \! \alpha \,(\gamma_\parallel - D \gamma_\parallel+1) \right]}{2 u_{x0}} +
\frac{\cos \! \alpha \, \sin^2 \! \alpha \, \beta_\bot (\gamma_\bot+D-3)}{2 u_{x0}} \bigg\} b_z \\ \nonumber
\frac{\mathrm{d} b_z}{\mathrm{d} x} & = & \left[ \frac{u_{x0}}{\cos \! \alpha}+\frac{\cos \! \alpha \,(\beta_\parallel-\beta_\bot-2)}{2 u_{x0}} \right] b_y.
\end{eqnarray}
Eqs. (\ref{eq:dbidx_lin_iso}) or (\ref{eq:dbidx_lin_anis}) can be generally rewritten as
\begin{eqnarray}
\label{eq:dbidx_general}
\frac{\mathrm{d} b_y}{\mathrm{d} x} = C b_z, \quad
\frac{\mathrm{d} b_z}{\mathrm{d} x} = E b_y.
\end{eqnarray}
Therefore using standard methods of analysis of two-dimensional dynamical
systems (linear and autonomous) we can investigate the behavior of the system
in the vicinity of the background state. Solutions with exponentially growing
amplitude are obtained for $CE>0$, otherwise we have oscillations
around the background state. The exponentially growing solutions have been
identified as solitary waves and the oscillations as linear
(small-amplitude) waves (see e.g. \cite{Sta04,Sta04b,Sta05,McKetal04}
where this kind of approach to wave solutions in fluid models of plasmas
has been extensively discussed).

\begin{figure}[!htbp]
\begin{center}
\includegraphics[scale=0.3]{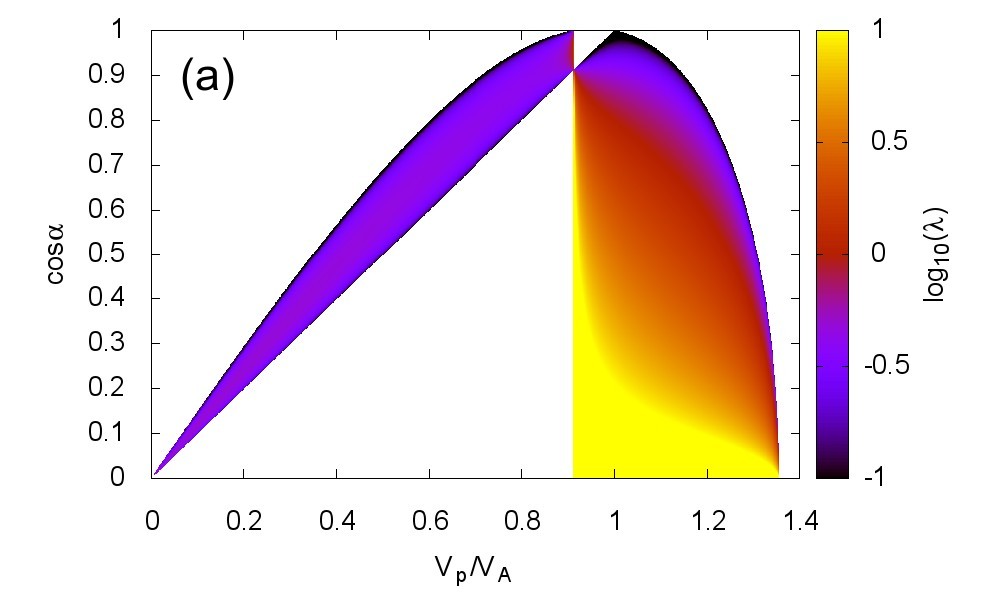}
\includegraphics[scale=0.3]{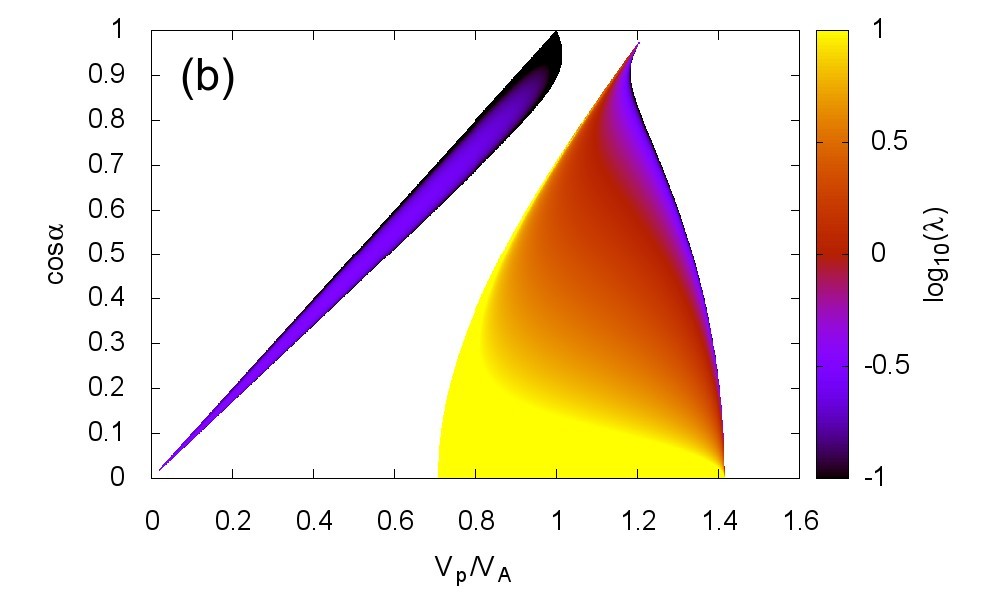}
\caption{Regions (shown in color) of possible existence of solitary solutions in
$V_p$--$\cos \alpha$ parameter plane for (a) isotropic and (b) anisotropic pressure.
The color scale corresponds to the growth rate $\lambda=\sqrt{C E}$
(only those regions are shown where $\lambda$ is real).}
\label{f_phplts}
\end{center}
\end{figure}

The solitary solutions in the HMHD model can be parameterized in terms of their
propagation speed $V_p=u_{x0}$ and the propagation angle $\alpha$ relative to the direction
of the background magnetic field. The analysis method outlined in the
previous paragraph indicates that solitary solutions (initially exponentially
growing with $x$) can be obtained only for a subset of the 
$V_{p}$--$\cos \alpha$ parameter plane as shown in
Fig. \ref{f_phplts}. In the colored regions the spatial
growth rate $\lambda=\sqrt{C E}$ is real, whereas
in the white regions it is imaginary (which corresponds to a linear wave regime).
Results of parametric analysis for the isotropic pressure
($\gamma=5/3$, $\beta_0=1$) are shown in Fig. \ref{f_phplts}(a)
and for the anisotropic pressure
($\gamma_\bot=2$, $\gamma_\parallel=3$, $\beta_{\bot 0}=\beta_{\parallel 0}=1$)
-- in Fig. \ref{f_phplts}(b).

\subsection{Nonlinear wave solutions of the HMHD equations as a tool for validation of numerical codes}

\begin{figure}[!htbp]
\begin{center}
\includegraphics[scale=0.35]{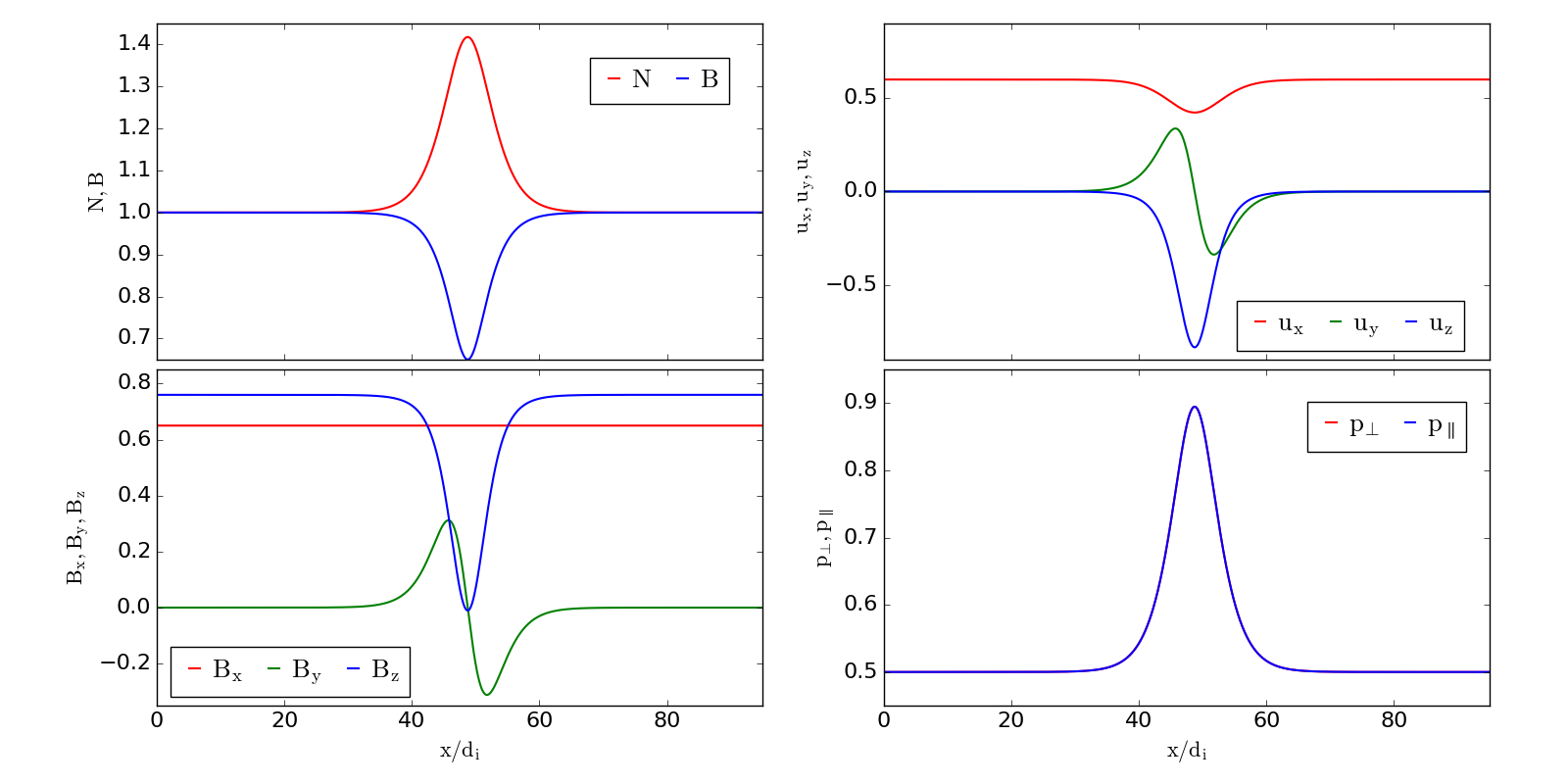}
\caption{An example of solitary wave for the isotropic pressure.}
\label{f_sol_iso}
\end{center}
\end{figure}

\begin{figure}[!htbp]
\begin{center}
\includegraphics[scale=0.35]{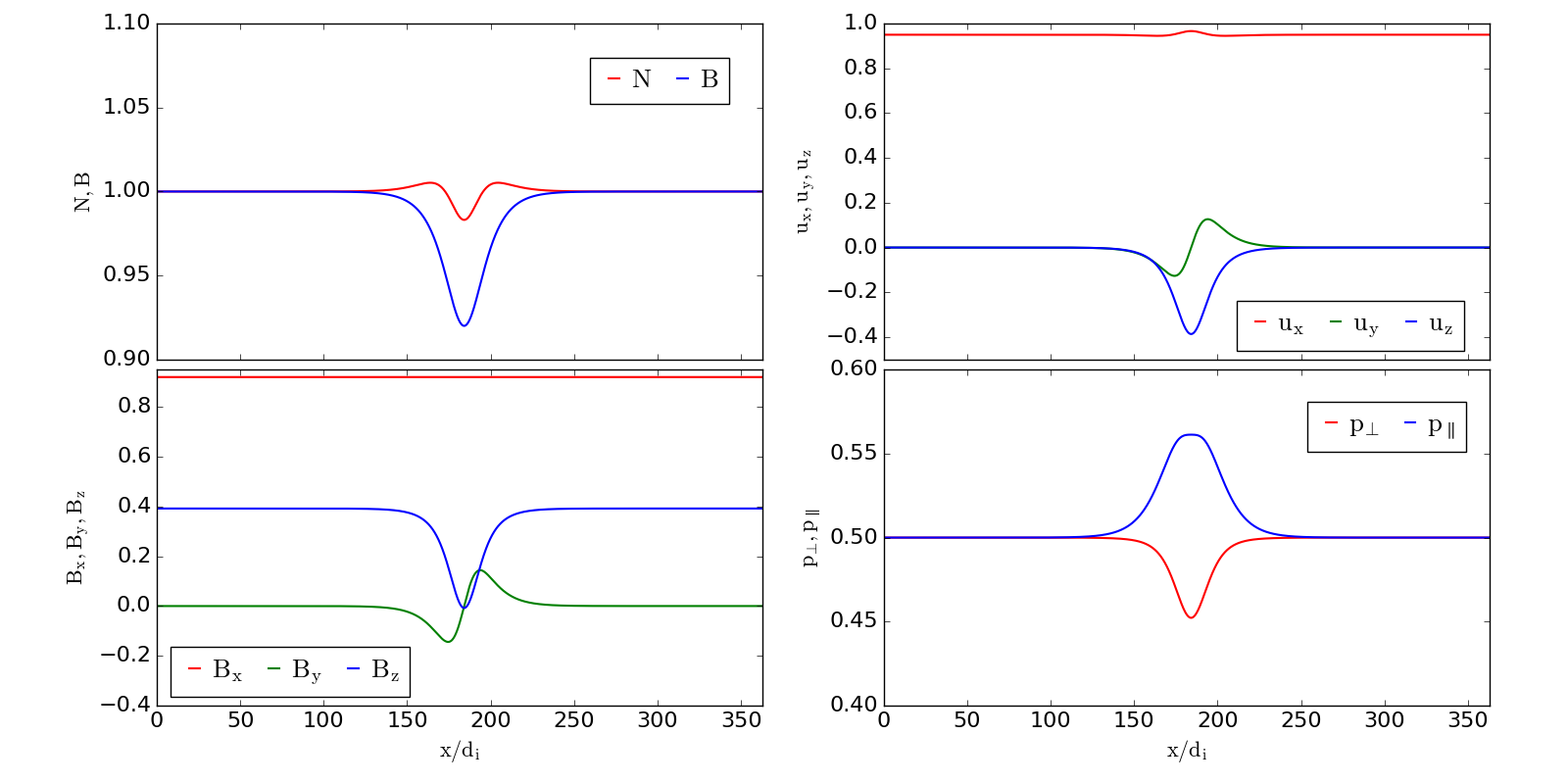}
\caption{An example of solitary wave for the anisotropic pressure.}
\label{f_sol_anis}
\end{center}
\end{figure}

Examples of the solitary wave solutions are shown in Fig. \ref{f_sol_iso}
(isotropic pressure, $\gamma=5/3$, $\beta_0=1$, $V_p=0.6$, $\cos \! \alpha=0.65$) and Fig.
\ref{f_sol_anis} (anisotropic pressure, CGL closure, $\gamma_\bot=2$, $\gamma_\parallel=3$, $\beta_{\bot 0}=\beta_{\parallel 0}=1$,
$V_p=0.95$, $\cos \! \alpha=0.92$). Spatial profiles of all fluid variables are
shown in the plots.

The solitary wave profiles can be used as the initial condition for fully time-dependent simulation. By
studying wave profile distortions resulting from propagation of
the waves using the numerical code one may obtain insight into properties of
the numerical algorithm, its resolution scaling properties and
correctness of the code implementation. For simplicity,
in this approach we use polytropic state equations instead of the
energy conservation equation  and the magnetic diffusivity is set to
zero. However, the conservation of
the total energy in the absence of heat fluxes and viscous/Joule's
heating leads to polytropic relations:
$p \propto N^{5/3}$ for the isotropic thermal pressure
and the CGL relations $p_\bot \propto N B$, $p_\parallel \propto N^3/B^2$
for the anisotropic gyrotropic pressure tensor.
Therefore the polytropic relations can be used to
obtain the nonlinear wave profiles that are useful for testing simulation codes with
the energy conservation equation explicitly included (which is sometimes
preferred to obtain better numerical stability and accuracy).

Eqs. (\ref{eq:dbidx}) give wave profiles in the
wave frame of reference, thus using them directly as the initial condition in
a simulation can be considered as a steady-state testing method.
In the plasma rest frame, the structures discussed above are seen as waves
propagating with the velocity $(V_p, 0, 0)$. Generally, by applying
the velocity transformation $u'_x=u_x-u_{x0}$ to the solitary solutions
described above it is possible to change
the wave frame to another frame where the wave propagates with the velocity $u_{x0}$ in the
simulation box. In particular, transformation to the plasma frame is obtained 
for $u_{x0}=V_p$. In Sec. \ref{sec:code_test} we present some examples
of the application of the nonlinear wave solutions for the validation of our
numerical code for HMHD simulations.

\section{Numerical scheme for time dependent simulations}

In general, the set of Eqs. (\ref{eq:dndtN})-(\ref{eq:dedtN}) can be considered as
\begin{equation}
\frac{\mathrm{d} \mathbf{U}}{\mathrm{d} t}=\mathbf{f}(\mathbf{U},t).
\end{equation}
where $\mathbf{U}(\mathbf{r},t)$ represents the state of the system
at a given time $t$ in a spatial location $\mathbf{r}$, and $\mathbf{f}$
is a nonlinear function that does not involve the time derivatives.
In the proposed numerical scheme the HMHD equations are advanced in time
using the second-order Runge-Kutta scheme
\begin{equation}
\mathbf{U}'=\mathbf{U}_n+ \mathbf{f}(\mathbf{U}_n) \, \Delta_t, \quad
\mathbf{U}''=\mathbf{U}'+\mathbf{f}(\mathbf{U}') \, \Delta_t , \quad 
\mathbf{U}_{n+1}=(\mathbf{U}_n+\mathbf{U}'')/2
\label{eq_rk2}
\end{equation}
describing the time evolution from the state $\mathbf{U}_n$ to $\mathbf{U}_{n+1}$,
where indexes $n$ and $n+1$ denote two subsequent time steps, and $\Delta_t$
is the integration time step.

The HMHD equations are solved on a uniform Cartesian grid. The
computational domain of the size $L_x \times L_y \times L_z$ is resolved
by $N_x \times N_y \times N_z$ cells, thus the spatial resolution of the grid is
$\Delta_x=L_x/N_x$, $\Delta_y=L_y/N_y$, $\Delta_z=L_z/N_z$.
The triplet of integers $i,j,k$ points at a cell center, $i$
numbers the cells in $x$ direction, $j$ along $y$, and $k$ along $z$. By using
$1/2$ in one of the indexes in the triplet we denote a face between cells in a given
direction, e.g. $i,j+1/2,k$ identifies the face between the cells $i,j,k$
and $i,j+1,k$ (a face normal to $y$ direction). When $1/2$ appears in two
indexes of the triplet, we refer to the edge between cells, e.g.
$i+1/2,j+1/2,k$ identifies the edge between the cells $(i,j,k)$,
$(i+1,j,k)$, $(i,j+1,k)$, $(i+1,j+1,k)$. One should note that it is a
standard notation used in literature (see, e.g. Refs.
\cite{Kunetal14,BalSpi99}, where plots illustrating the notation are
presented).

All the transport equations except for Eq. (\ref{eq:dbdtN}) (magnetic field
transport) are numerically solved using the Kurganov-Tadmor scheme \cite{KurTad00}
from a family of MUSCL schemes based on linear piecewise approximation for every computational cell.
For those equations $f_q(\mathbf{U})$ ($q = {x, y, z}$) from Eq. (\ref{eq_rk2}) is approximated as
\begin{equation}
f_q(\mathbf{U}) \approx \frac{F_{i + 1/2,j,k}-F_{i - 1/2,j,k}}{\Delta_x} +
\frac{F_{i,j + 1/2,k}-F_{i,j - 1/2,k}}{\Delta_y}+\frac{F_{i,j,k+1/2}-F_{i,j,k-1/2}}{\Delta_z}
\end{equation}
As suggested by indexes containing $1/2$, the fluxes $F$ are computed at the cell faces.
One should also note that in the first term on the right-hand side the fluxes are
computed at the faces normal to $x$ direction, in the second -- normal to $y$,
in the third -- normal to $z$.

The numerical fluxes can be written as the sum $F_{i \pm 1/2,j,k}=F^C_{i \pm 1/2,j,k}+F^H_{i \pm 1/2,j,k}$,
where $F^C_{i \pm 1/2,j,k}$ represents the Rusanov flux for the MHD part of the equations
(i.e. without the Hall term) and $F^H_{i \pm 1/2,j,k}$ contains
Hall corrections. The classical MHD flux is defined as
\begin{equation}
F^C_{i \pm 1/2,j,k} = \frac{1}{2} \left[ F^C \left( u^R_{i \pm 1/2,j,k} \right) + F^C \left( u^L_{i \pm 1/2,j,k} \right) \right] -
\frac{1}{2} c_{i \pm 1/2,j,k} \left[ u^R_{i \pm 1/2,j,k} - u^L_{i \pm 1/2,j,k} \right]
\end{equation}
where the local propagation speed $c_{i \pm 1/2,j,k}$ is the maximum absolute eigenvalue of the Jacobian
of $F^C$ over cells $i, i \pm 1$. The left (L) and right (R) states are computed as
\begin{eqnarray}
u^L_{i - 1/2,j,k}=u_{i-1,j,k}+\frac{1}{2} \phi(r_{i-1,j,k}) (u_{i,j,k}-u_{i-1,j,k}), \quad
u^R_{i - 1/2,j,k}=u_{i,j,k}-\frac{1}{2} \phi(r_{i,j,k}) (u_{i+1,j,k}-u_{i,j,k}) \nonumber \\
u^L_{i + 1/2,j,k}=u_{i,j,k}+\frac{1}{2} \phi(r_{i,j,k}) (u_{i+1,j,k}-u_{i,j,k}), \quad
u^R_{i + 1/2,j,k}=u_{i+1,j,k}-\frac{1}{2} \phi(r_{i+1,j,k}) (u_{i+2,j,k}-u_{i+1,j,k})
\label{eq_rlstates}
\end{eqnarray}
where $r_{i,j,k}=(u_{i,j,k}-u_{i-1,j,k})/(u_{i+1,j,k}-u_{i,j,k}+\epsilon_E)$ and
the van Leer flux limiter $\phi(r)=(r+|r|)/(1+|r|)$ is used.
One should note that Eqs. (\ref{eq_rlstates}) are applied to
primitive variables, that are then used to compute conservative variables
and finally the fluxes. The fluxes $F^C_{i,j \pm 1/2,k}$ and $F^C_{i,j,k \pm 1/2}$
at the faces normal to $y$ and $z$ directions
can be defined analogously to the flux at the face normal to
$x$ direction $F^C_{i \pm 1/2,j,k}$ by changing only the leading dimension in the above definitions.

The Hall corrections $F^H_{i \pm 1/2,j,k}$ to the fluxes are
computed separately using the averaged variables $u_{i \pm
1/2,j,k}=(u^L_{i \pm 1/2,j,k}+u^R_{i \pm 1/2,j,k})/2$. To compute the current
density $\mathbf{J} \propto \nabla \times B$ included in the Hall
velocity $\mathbf{U}_H \propto -\mathbf{J}/N$ we need the spatial
derivatives of the magnetic field components. In the scheme proposed here,
the normal derivatives are computed in a different manner than the
tangential derivatives. This approach is similar to that presented in
Ref. \cite{Totetal08}, but we use additional averaging of the normal
derivatives. The explicit form of the current density components for the
faces normal to the $x$ direction is
\begin{eqnarray}
J^x_{i+1/2,j,k} & = &
\frac{B^z_{i,j+1,k}+B^z_{i+1,j+1,k}-B^z_{i,j-1,k}-B^z_{i+1,j-1,k}}{4 \Delta_y} -
\frac{B^y_{i,j,k+1}+B^y_{i+1,j,k+1}-B^y_{i,j,k-1}-B^y_{i+1,j,k-1}}{4 \Delta_z}, \nonumber \\
J^y_{i+1/2,j,k} & = &
\frac{B^x_{i,j,k+1}+B^x_{i+1,j,k+1}-B^x_{i,j,k-1}-B^x_{i+1,j,k-1}}{4 \Delta_z} -
\frac{B^z_{i+1,j,k}-B^z_{i,j,k}}{3 \Delta_x} - \nonumber \\
& & \frac{B^z_{i+1,j-1,k}-B^z_{i,j-1,k}}{6 \Delta_x} - \frac{B^z_{i+1,j+1,k}-B^z_{i,j+1,k}}{6 \Delta_x} -
\frac{B^z_{i+1,j,k-1}-B^z_{i,j,k-1}}{6 \Delta_x} - \frac{B^z_{i+1,j,k+1}-B^z_{i,j,k+1}}{6 \Delta_x}, \nonumber \\
J^z_{i+1/2,j,k} & = & -
\frac{B^x_{i,j+1,k}+B^x_{i+1,j+1,k}-B^x_{i,j-1,k}-B^x_{i+1,j-1,k}}{4 \Delta_y} +
\frac{B^y_{i+1,j,k}-B^y_{i,j,k}}{3 \Delta_x} + \nonumber \\
& & \frac{B^y_{i+1,j-1,k}-B^y_{i,j-1,k}}{6 \Delta_x} + \frac{B^y_{i+1,j+1,k}-B^y_{i,j+1,k}}{6 \Delta_x} +
\frac{B^y_{i+1,j,k-1}-B^y_{i,j,k-1}}{6 \Delta_x} + \frac{B^y_{i+1,j,k+1}-B^y_{i,j,k+1}}{6 \Delta_x},
\end{eqnarray}
which can be compared with Eq. (16) in Ref. \cite{Totetal08} to illustrate the
differences. The tangential derivatives are computed by central
differencing and averaging in the $i$ direction (face normal direction).
In fact, the normal derivatives can be also considered as obtained by
central differencing (note that the value of the derivative at $i+1/2$ is
needed), but they are averaged in $j$ and $k$ directions with different
weights for the central point $j,k$ and neighboring points $j \pm 1,k \pm
1$. Using the current density vector we can compute the Hall corrections
$F^H_{i \pm 1/2,j,k}$. The current density components (and Hall corrections)
for the faces normal to the $y$ and $z$ directions can be defined analogously, according
to the rules described above.

The equation of the magnetic field transport is advanced in time by using so-called
flux-interpolated constrained-transport (flux-CT) approach \cite{BalSpi99}, that was found
to be one of the most accurate in a series of tests reported in Ref. \cite{Tot00}.
The flux-CT approach was developed on the basis of a specific discretization scheme referred
often to as a staggered-mesh transport algorithm for the magnetic field
(see e.g. \cite{Yee66,EvaHaw88}).
In this approach, a discrete version of the Stokes' theorem is used for updating magnetic
field components in time. Different magnetic field components are collocated on different
cell faces (normal to a given component). On the other hand, the electric field components are collocated at
the edges of the cells. A version of this algorithm proposed in Ref. \cite{BalSpi99} uses
a duality between the electric field and the fluxes that determine the
transport of the magnetic field. Assuming that
$F^{B_p}_{q;i \pm 1/2,j,k}=F^{B_p;C}_{q;i \pm 1/2,j,k}+F^{B_p;H}_{q;i \pm 1/2,j,k}$ is the
component of the flux normal to the $q$ direction in the equation of the transport
of the $p$ component of the magnetic field, the electric field components at the cell edges are
\begin{eqnarray}
E_{x;i,j+1/2,k+1/2}= \frac{1}{4} \left( F^{B_y}_{z;i,j,k+1/2} + F^{B_y}_{z;i,j+1,k+1/2} -
F^{B_z}_{y;i,j+1/2,k} - F^{B_z}_{y;i,j+1/2,k+1} \right), \nonumber \\
E_{y;i+1/2,j,k+1/2}= \frac{1}{4} \left( F^{B_z}_{x;i+1/2,j,k} + F^{B_z}_{x;i+1/2,j,k+1} -
F^{B_x}_{z;i,j,k+1/2} - F^{B_x}_{z;i+1,j,k+1/2} \right), \nonumber \\
E_{z;i+1/2,j+1/2,k}= \frac{1}{4} \left( F^{B_x}_{y;i,j+1/2,k} + F^{B_x}_{y;i+1,j+1/2,k} -
F^{B_y}_{x;i+1/2,j,k} - F^{B_y}_{x;i+1/2,j+1,k} \right).
\end{eqnarray}
The advancing of the magnetic field components in time is done by using the following scheme
\begin{eqnarray}
B^{n+1}_{x;i+1/2,j,k}=B^{n}_{x;i+1/2,j,k}-\Delta_t \left[ \frac{E_{z;i+1/2,j+1/2,k}-E_{z;i+1/2,j-1/2,k}}{\Delta_y}
- \frac{E_{y;i+1/2,j,k+1/2}-E_{y;i+1/2,j,k-1/2}}{\Delta_z} \right], \nonumber \\
B^{n+1}_{y;i,j+1/2,k}=B^{n}_{y;i,j+1/2,k}-\Delta_t \left[ \frac{E_{x;i,j+1/2,k+1/2}-E_{x;i,j+1/2,k-1/2}}{\Delta_z}
- \frac{E_{z;i+1/2,j+1/2,k}-E_{z;i-1/2,j+1/2,k}}{\Delta_x} \right], \nonumber \\
B^{n+1}_{z;i,j,k+1/2}=B^{n}_{z;i,j,k+1/2}+\Delta_t \left[ \frac{E_{x;i,j+1/2,k+1/2}-E_{x;i,j-1/2,k+1/2}}{\Delta_y}
- \frac{E_{y;i+1/2,j,k+1/2}-E_{y;i-1/2,j,k+1/2}}{\Delta_x} \right].
\end{eqnarray}
The values of the magnetic field components in the cell centers are computed as
two-point averages
\begin{equation}
B^{n+1}_{x;i,j,k}=\frac{B^{n+1}_{x;i+1/2,j,k}+B^{n+1}_{x;i-1/2,j,k}}{2}, \quad
B^{n+1}_{y;i,j,k}=\frac{B^{n+1}_{y;i,j+1/2,k}+B^{n+1}_{y;i,j-1/2,k}}{2}, \quad
B^{n+1}_{z;i,j,k}=\frac{B^{n+1}_{z;i,j,k+1/2}+B^{n+1}_{y;i,j,k-1/2}}{2}.
\end{equation}
If $\nabla \cdot \mathbf{B}=0$ in the initial condition, the flux-CT scheme maintains
the solenoidal magnetic field during the time evolution to the accuracy comparable with
the machine round off error.

The resistive terms in Eqs. (\ref{eq:dbdtN}) and (\ref{eq:dedtN}) also require the estimation
of the current density components by differencing the magnetic field components. The
same rule as in the case of the Hall-term corrections is used here, i.e. the normal derivatives
use two nearest cells only, whereas the tangential derivatives are computed by central
differencing and averaging in the normal direction.

The HMHD model is known to include dispersive whistler waves that impose
strong constraints on the time step of the explicit numerical scheme presented
above. To ensure the stability of the scheme we use
$\Delta_t=C (\Delta_x/c_{wx}+\Delta_y/c_{wy}+\Delta_z/c_{wz})$,
where the fastest wave speed in $i$ direction is estimated to be
$c_{wi}=|u_i|+c_{fi}+\frac{2 B \pi}{N \Delta_i}$,
$c_f$ is the fast magnetosonic speed \cite{Hub03}. In
the simulations discussed below we assume the Courant number $C=0.4$.

\section{Implementation summary}
The code was implemented in C/C++ in a modular manner with separate
procedures for setting up problem-specific initial condition.
Both periodic and floating (zero normal gradient)
boundary conditions were implemented. The
boundary conditions are changed at the compilation time by setting
appropriate preprocessor directives during compilation. The simulation box is decomposed into
smaller boxes depending on the number of available computational cores.
The MPI library is used for the exchange of information between the cores
about their boundaries.  Simulations in one-, two- and
three-dimensional simulation box are possible by setting appropriate preprocessor
directives at the compilation time.

\section{Numerical tests}
\label{sec:code_test}

\subsection{Nonlinear solitary waves}

The solitary solutions of the HMHD equations discussed in Sec. \ref{sec:solit}
give a possibility of verifying the correctness of the implementation and
testing the accuracy of the numerical simulations. Since the amplitude of the
fluctuations of the magnetic field components is not small
as compared with the mean magnitude of the magnetic field (see, e.g. Figs. \ref{f_sol_iso}
and \ref{f_sol_anis}), this
testing method can be considered as a validation in the nonlinear regime. The exact analytical form for HMHD
solitary solutions is not known in general, but solitary waves can be obtained as a numerical
solution of the set of ordinary differential equations (\ref{eq:dbidx}). In its own
frame of reference a solitary wave is a steady-state solution. We can easily
change it into a propagating solution by simply adding a constant $u_{x0}$ to
velocity component $u_x$ of the solitary wave profile. By rotating the
structure (and all vector quantities) in a two- or three-dimensional simulation box
we can test oblique propagation with respect to the grid.
In tests presented in this subsection the magnetic diffusivity is $\eta=0$.

One should note that using solitary
solutions for testing the time-dependent simulations imposes strong
requirements on the accuracy of the solitary solution itself. In the
discussion presented below we used a procedure \verb+odeint()+ from
Ref. \cite{Preetal92} with an accuracy parameter $\mathrm{eps}=10^{-12}$ for solving
Eq. (\ref{eq:dbidx}) and a procedure \verb+rtbis()+ with a parameter $\mathrm{xacc}=10^{-16}$
for Eqs. (\ref{eq:nb}) or (\ref{eq:nb_anis}). In the testing procedure reported below,
one solution with 16384 points is obtained by integration of ODE and then it
is used as the initial condition in simulations with the resolution $N_x$
by removing every $16384/N_x$ points. It is also used as a base
solution for oblique propagation with respect to the grid in 2D simulations as discussed below.

The first test verifies the scheme properties for a steady-state
solution. Solitary waves shown in Figs. \ref{f_sol_iso} (isotropic pressure) and
\ref{f_sol_anis} (anisotropic pressure) are used as the initial
\begin{figure}[!htbp]
\begin{center}
\includegraphics[scale=0.3]{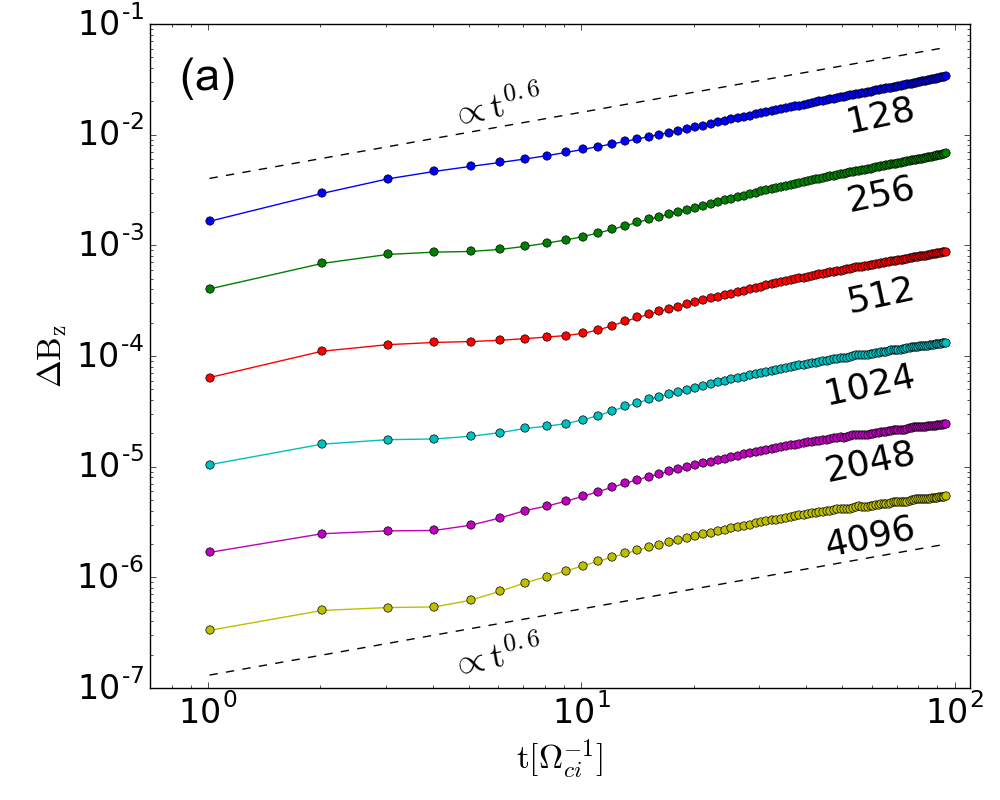}
\includegraphics[scale=0.3]{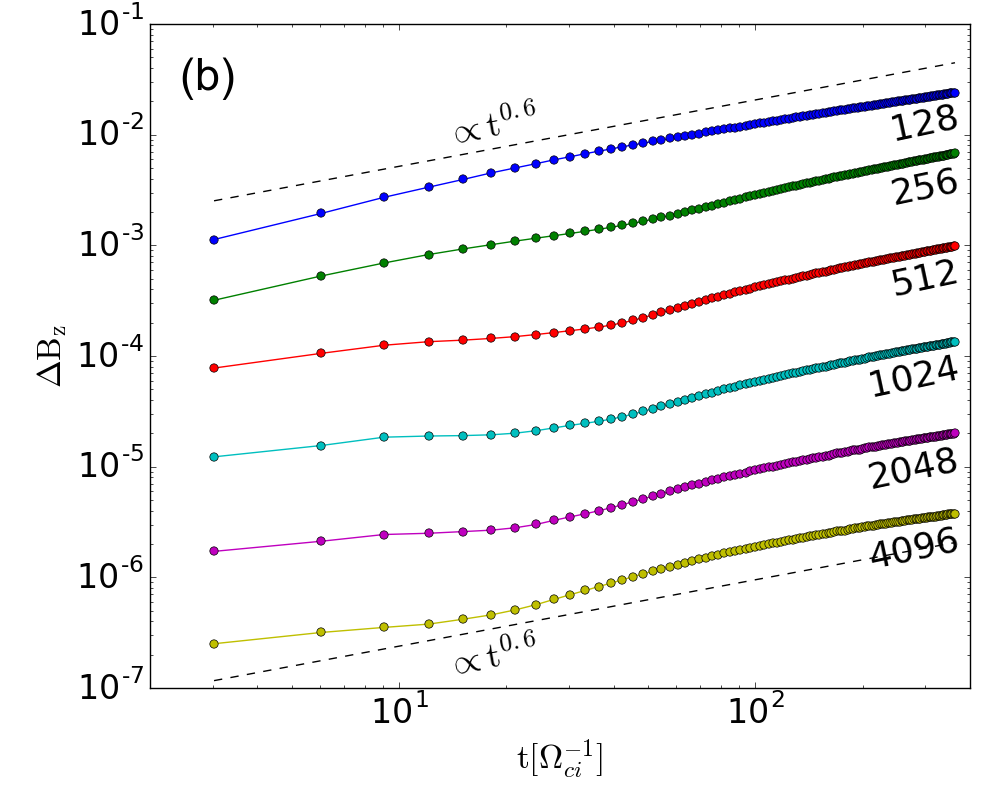}
\caption{Time dependence of the error in a steady-state test
with a solitary wave for (a) isotropic and (b) anisotropic pressure case.
Results for six different resolutions are presented $N_x=$ 128,
256, 512, 1024, 2048, 4096. Dashed lines show $\propto t^{0.6}$
scaling that suggests nearly diffusive character of the growth of errors in time.}
\label{f_err_steady}
\end{center}
\end{figure}
condition. The simulation is done in the frame of the wave, thus we can
check the accuracy of the code for maintaining a steady-state solution in
the one-dimensional simulation. Periodic boundary conditions are used.
For the isotropic case the size of the simulation domain is $L_x=95 \, d_\mathrm{i}$,
for the anisotropic case $L_x=363 \, d_\mathrm{i}$. Six resolutions are tested: $N_x=$128,
256, 512, 1024, 2048, 4096 grid points, the final simulation time is
$t_\mathrm{max}=L_x/V_A=95$ $\Omega_{i}^{-1}$ for the isotropic case
and $t_\mathrm{max}=L_x/V_A=363$ $\Omega_{i}^{-1}$ for the
anisotropic case. Fig. \ref{f_err_steady} shows the time
dependence of the error of the transversal component of the magnetic field $\Delta
B_z(t)=\sqrt{\sum_{i=1}^{N_x}[B_{z,i}(t)-B_{z,i}(0)]^2}$ that increases with time as
$\Delta B_z(t) \propto t^{0.6}$ indicating nearly diffusive character of errors
introduced by the numerical code. Resolution change of factor $\sim$2 gives
$\sim$4-fold smaller error that confirms the expected second-order scaling
of the numerical scheme under verification.

\begin{figure}[!htbp]
\begin{center}
\includegraphics[scale=0.3]{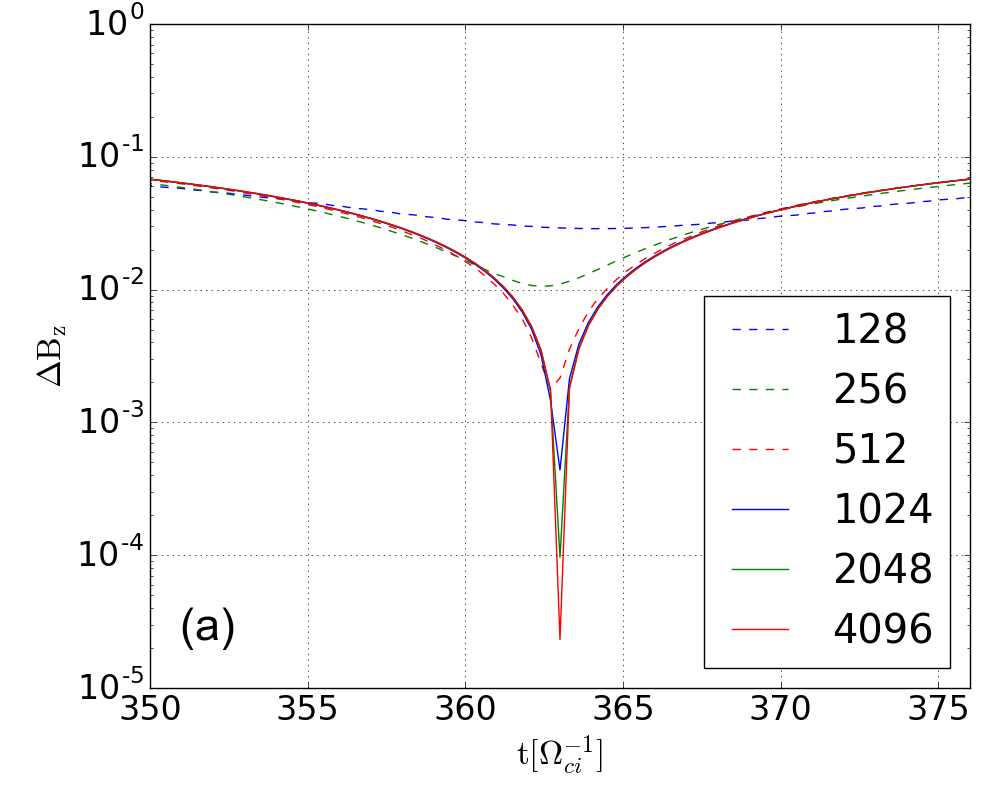}
\includegraphics[scale=0.3]{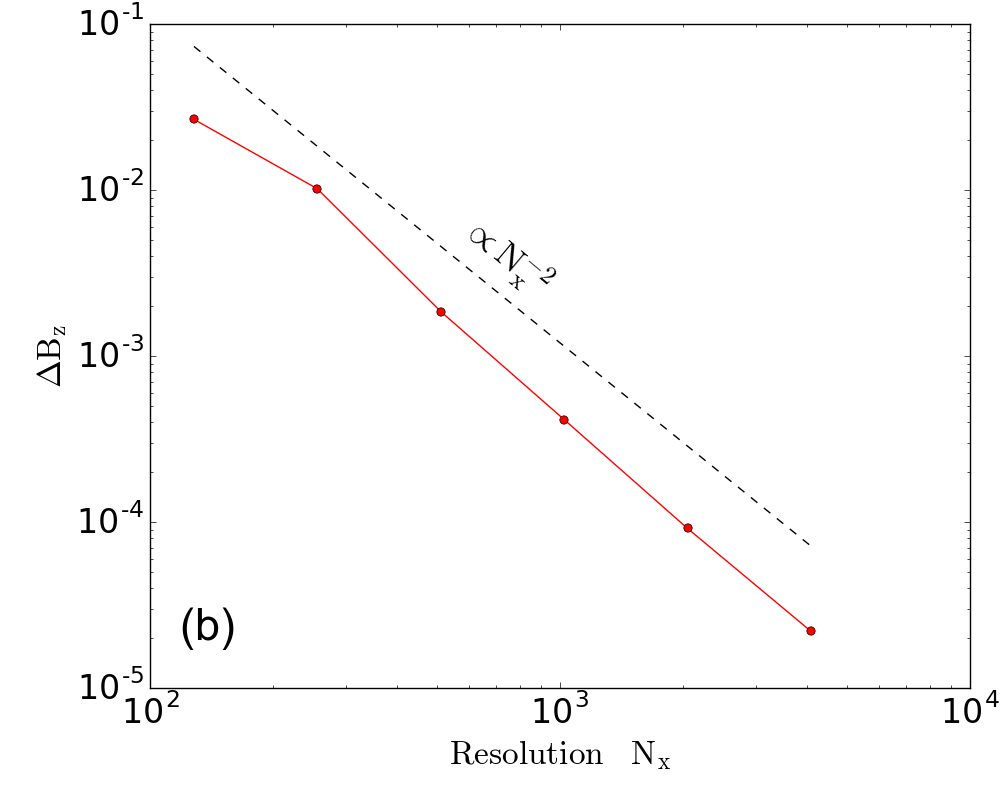}
\caption{Time dependence of the error in a propagating-solitary-wave test (panel (a) on the left).
Results for six different resolutions are presented $N_x=$ 128,
256, 512, 1024, 2048, 4096. Local minima of $\Delta B_z(t)$ are
related to arriving the wave to its initial position.
Panel (b) shows the dependence of the local minima on
the resolution $N_x$. Dashed black line corresponds to the expected
$\propto N_x^{-2}$ scaling.}
\label{f_err_prop}
\end{center}
\end{figure}

In the second test, the same solitary solution is amended by adding a
constant propagation velocity $u_{x0}=1.0$ (it corresponds to the
propagation in the simulation box frame with the Alfven speed), which makes it possible
to test the code for a propagating structure. Periodic boundary
conditions are applied and for the anisotropic pressure case shown in Fig.
\ref{f_sol_anis}, the size of the simulation
domain is $L_x=363$ $d_i$. Therefore after the time $t'=L_x/u_{x0}=363$ $\Omega_{i}^{-1}$ the wave should
arrive back at its initial position. Fig. \ref{f_err_prop}(a) shows dependence
of $\Delta B_z(t)$ for this type of test for six different resolutions.
Since the amplitude and velocity of solitary waves are related, the amplitude
decrease related to errors introduced by the numerical scheme changes the propagation
speed of the wave with respect to plasma frame. This
is clearly seen for low resolutions as a shift between the time of the local minimum of $\Delta B_z(t)$ and $t'$.
As the resolution increases the shift becomes smaller. In Fig. \ref{f_err_prop}(b) we show
the minimum of $\Delta B_z(t)$ from Fig. \ref{f_err_prop}(a) as a function of
the resolution $N_x$, which confirms clearly the second-order
scaling of the numerical errors $\Delta B_z(t) \propto N_x^{-2}$.
Fig. \ref{f_err_prop} shows results for the anisotropic-pressure case, a
similar study for the isotropic pressure gives the same scaling of numerical errors
(not shown here).

\begin{figure}[!htbp]
\begin{center}
\includegraphics[scale=0.3]{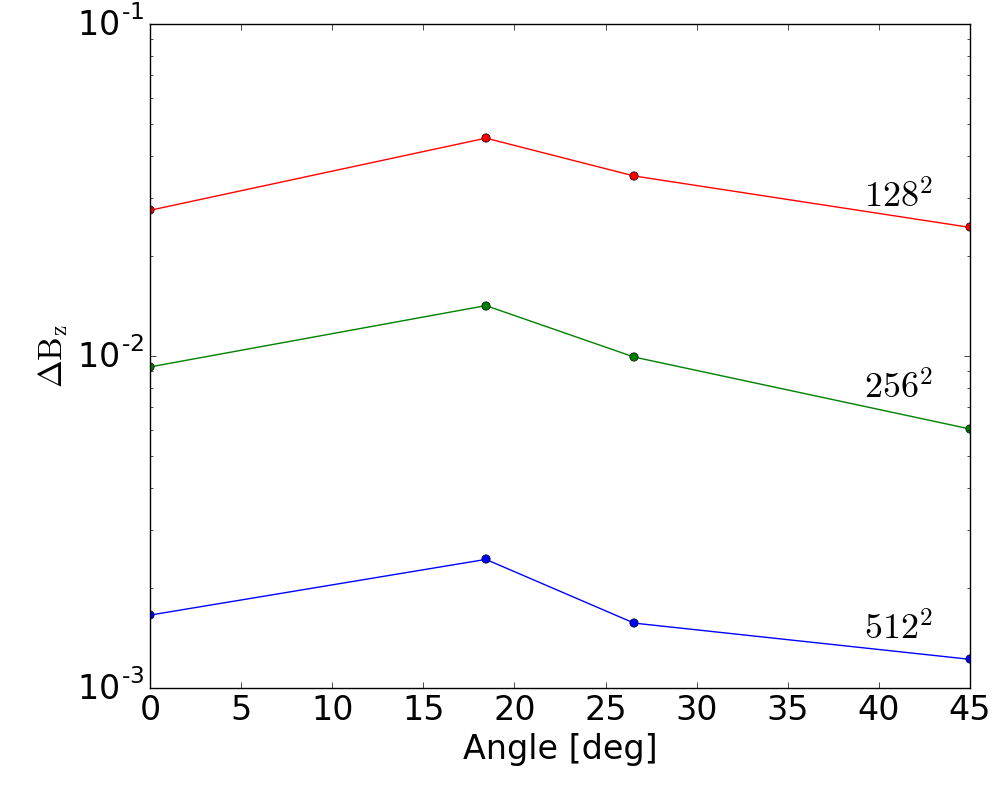}
\caption{Dependence of the error in a test with propagating structure in 2D simulation box
for different angles between the wave normal direction and $x$ direction (planar
solitary wave propagates in 2D simulation box at different angles to the simulation grid).}
\label{f_err_prop_oblique}
\end{center}
\end{figure}

Fig. \ref{f_err_prop_oblique} summarizes results of testing of oblique propagation
\begin{figure}[!htbp]
\begin{center}
\includegraphics[scale=0.3]{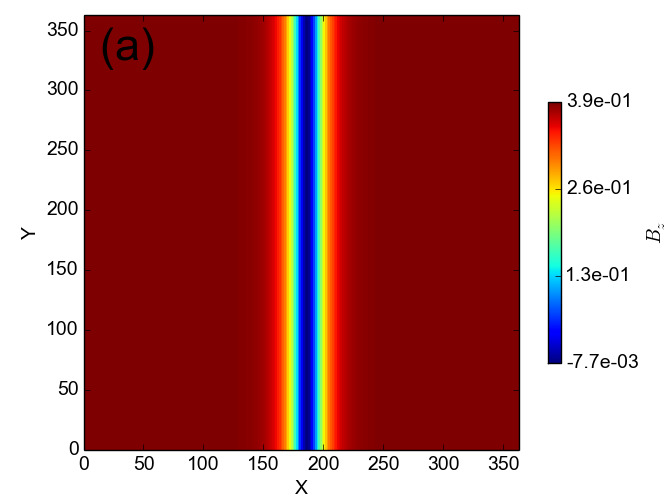}
\includegraphics[scale=0.3]{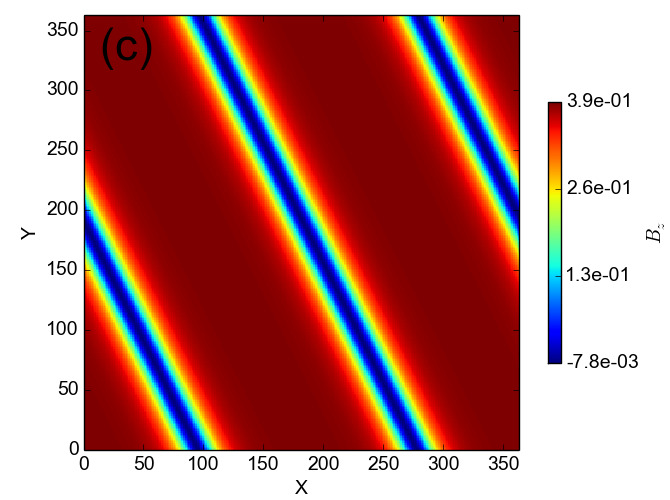}
\includegraphics[scale=0.3]{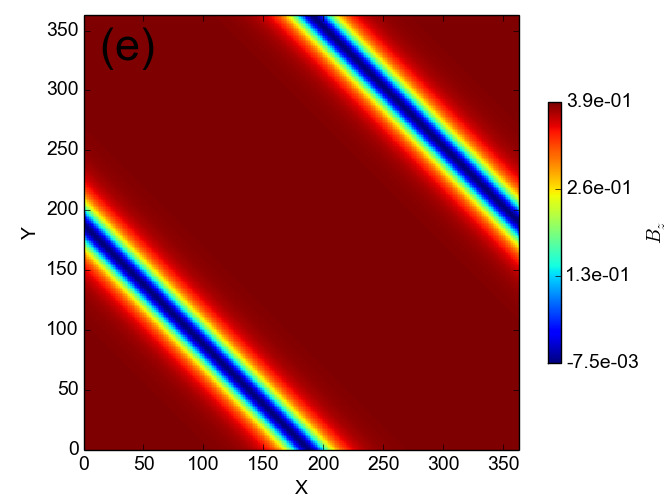}
\includegraphics[scale=0.3]{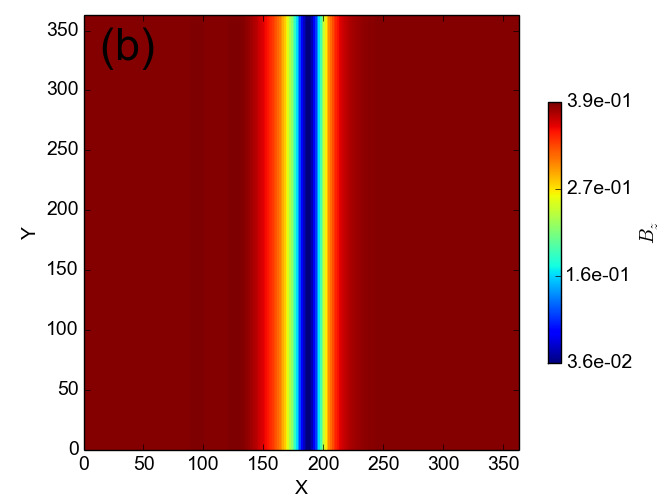}
\includegraphics[scale=0.3]{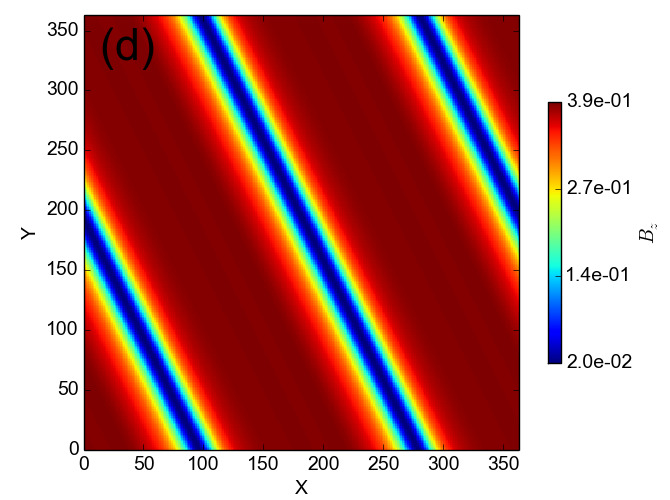}
\includegraphics[scale=0.3]{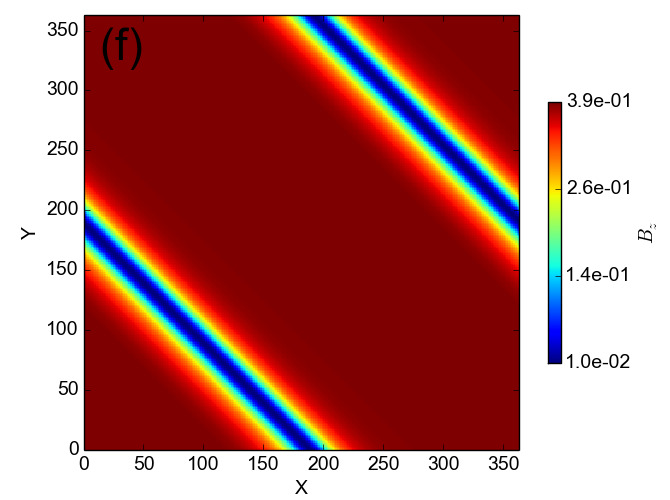}
\caption{Comparison of the initial condition at $t_0=0$ [panels (a),(c),(e)] and a later stage of evolution
at $t_1=(L_x/u_{x0}) \cos (\alpha)$ [panels (b) $t_1=363 \, \Omega_i^{-1}$, (d) $t_1=324.68 \, \Omega_i^{-1}$, (f) $t_1= 256.68 \, \Omega_i^{-1}$]
for oblique propagation test.
The propagation angle $\alpha$ (between wave normal direction and $x$ direction) is (a),(b)
0, (c),(d) 26.6, (e),(f) 45 degrees, correspondingly. The spatial resolution for the test is $N_x \times N_y = 256^2$.}
\label{f_err_prop_oblique_2}
\end{center}
\end{figure}
of a solitary wave in 2D simulation box for different angles between the wave normal
direction and $x$ direction. The wave propagates in the simulation box with
the velocity $u_{x0}=1.0$, similarly to 1D propagation test, the size of
the simulation domain is $L_x \times L_y = 363 \times 363 d_\mathrm{i}$.
Results for three resolutions are shown: $N_x \times N_y =$ $128^2$, $256^2$,
$512^2$ for the anisotropic-pressure solitary wave shown in Fig. \ref{f_sol_anis}.
Periodic boundary conditions are applied in $x$ and $y$ directions. Fig. \ref{f_err_prop_oblique}
suggests a weak dependence of the errors introduced by the numerical scheme on the
propagation angle with respect to the computational grid. As
illustrated in Fig. \ref{f_err_prop_oblique_2} numerical errors mainly
contribute to a decrease of the wave amplitude, which can be seen by comparison
of the range in the color bars in panels (a),(c),(e) with respect to (b),(d),(f).
The numerical errors also introduce asymmetry of the leading edge with respect to the trailing
edge (the wave propagates to the right, thus leading edge is on the right and trailing edge
on the left), which is particularly well seen by comparison of panels (a) and (b).
Similar results have been obtained for the isotropic case (not shown here).

\begin{figure}[!htbp]
\begin{center}
\includegraphics[scale=0.22]{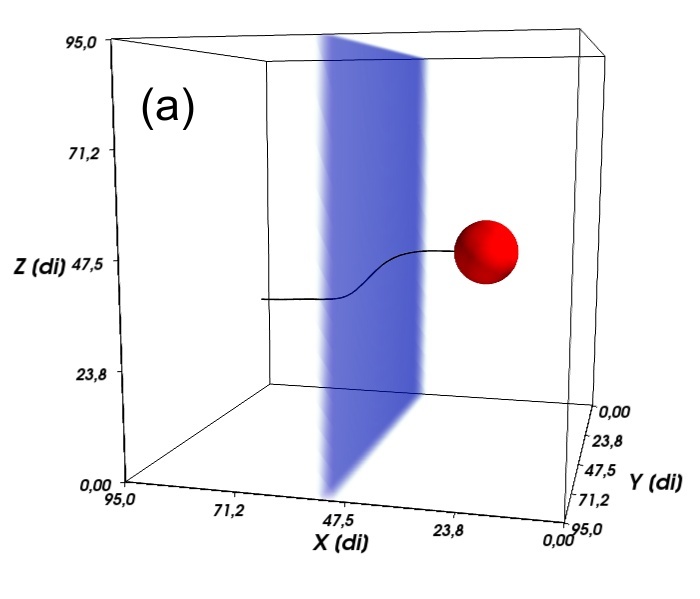}
\includegraphics[scale=0.22]{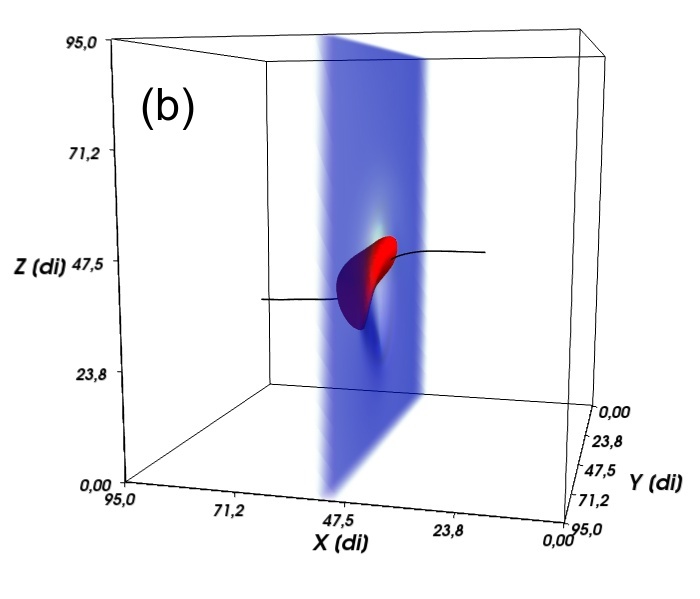}
\includegraphics[scale=0.22]{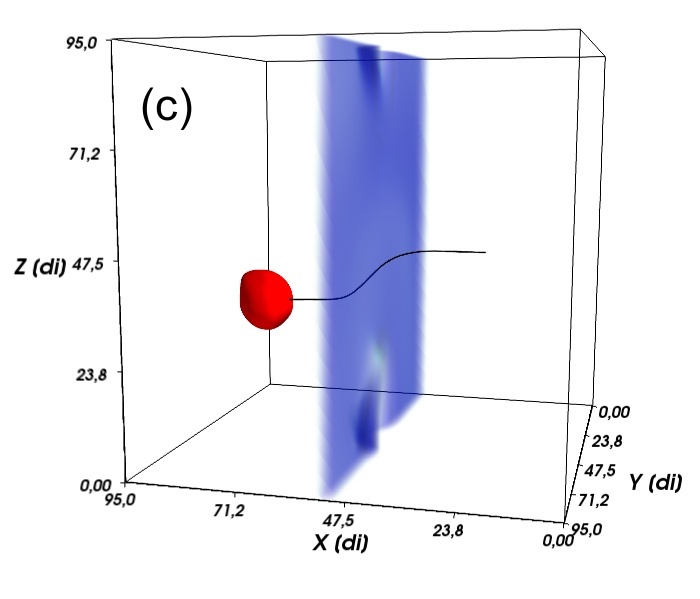}
\caption{
Three dimensional simulation of the interaction of a localized spherically-symmetric density enhancement
(blob, red color) with a planar solitary wave (blue planar structure). The blob pierces
the planar soliton that leads to appearance of a
perturbation after the interaction. The soliton puncture is only a transient
effect, post-interaction dynamical processes appear to work towards rebuilding the
soliton structure in its pierced part. Three frames are shown: (a) the initial
condition at $t=0 \, \Omega_i^{-1}$, (b) a moment of the interaction of the blob and soliton
at $t=52.5 \, \Omega_i^{-1}$, (c) a post-interaction state at $t=97.5 \, \Omega_i^{-1}$.
Black line shows approximately the trajectory of the center of the blob,
the blob itself is visualized by $N=1.75$ isosurface.
The planar soliton is shown using a volume
rendering technique in Mayavi visualization software \cite{RamVar11}, where the opacity
(non-zero for $0.7<B<0.8$) is controlled by a transfer
function properly adjusted to visualize the planar soliton.}
\label{f_err_solit_blob}
\end{center}
\end{figure}

Another test is intended to check the behavior of the code in the case
of strongly nonlinear localized interactions. The simulation
domain is 3D, the grid resolution is $N_x \times N_y \times N_z = 128^3$,
the size is $L_x = L_y = L_z = 95 \, d_i$,
periodic boundary conditions are used in all directions. In the initial
condition the solitary wave from Fig. \ref{f_sol_iso} (isotropic
pressure) is set up in the middle of the box (blue planar structure in Fig. \ref{f_err_solit_blob}(a)).
Additionally a localized spherically-symmetric density enhancement (blob) is set up
(the center of the blob is initially located at $x=23.75$ $d_i$, $y=z=47.5$ $d_i$).
The simulation is done in the soliton frame, thus the blob (that can be considered as an
entropy wave perturbation) is advected by
the flow towards the planar soliton. Fig. \ref{f_err_solit_blob}(b) shows
a moment of the interaction of the blob and the solitary wave, when the structures are being
distorted by the interaction process. During the interaction, the blob
is constantly blown downward (towards $z=0$) by the flow inside the solitary wave ($u_z<0$ in the solitary
wave as seen in Fig. \ref{f_sol_iso}). After having pierced the solitary wave, the blob
is advected by the flow towards the boundary as seen in Fig. \ref{f_err_solit_blob}(c).
The interaction process produces a perturbation in the planar soliton that is
advected downward by the flow inside the solitary wave. The perturbation is
seen at the bottom and at the top of the simulation box in Fig. \ref{f_err_solit_blob}(c)
due to periodic boundary conditions applied in the $z$ direction.

\subsection{Magnetic reconnection}
HMHD codes are conventionally tested with a well-studied (also by MHD and
kinetic codes) problem, the Geospace Environmental Modeling (GEM)
magnetic reconnection challenge \cite{Biretal01}. The exact form of the
solution is not known in this case, but solutions obtained using our code
can be compared with a number of solutions published elsewhere.
One should note that the problem of magnetic reconnection
is quite specific as it involves dynamics in regions, where particles are
weakly magnetized or unmagnetized. For these ``diffusive'' regions the
isotropic and gyrotropic models presented in our paper are not good
approximations since non-gyrotropic pressure tensor should be considered
to describe properly the underlying physics \cite{Wanetal15}.
Even though HMHD description gives a simplified picture, by including the Hall term we
can obtain reconnection rate much larger that in resistive MHD and in
some cases comparable with kinetic description. This kind of approach can
be useful for some problems of interest, where one considers consequences
of fast magnetic reconnection on large-scale dynamics, rather than details
of the diffusion-region physics. In any case, the GEM reconnection challenge 
has become a standard benchmark showing the correctness of implementation and reliability
of the applied numerical approach, therefore we present below tests for this problem.

In our simulation the problem is formulated in $x$--$y$ plane. The initial condition is
\begin{eqnarray}
B_x & = & B_0 \tanh[(y-y_0)/\lambda] + (\psi_0 \pi/L_y) \cos(2 \pi x/L_x) \sin[\pi (y-y_0)/L_y], \nonumber \\
B_y & = & - (2 \psi_0 \pi/L_x) \sin(2 \pi x/L_x) \cos[\pi (y-y_0)/L_y], \nonumber \\
N & = & N_0 \mathrm{sech}^2[(y-y_0)/\lambda]+N_\infty, \nonumber \\
P & = & 0.6- (B_0^2/2) \tanh[(y-y_0)/\lambda],
\end{eqnarray}
where $B_0=1$, $N_0=1$, $N_\infty=0.2$, $\psi_0=0.1$, $\lambda=0.5$, $y_0=6.4$. The simulation
domain size is $L_x=25.6$, $L_y=12.8$. The spatial resolution is $N_x \times N_y = 128^2$ points. In the $x$ direction
periodic boundary conditions are applied, in the $y$ direction we use zero-gradient floating boundary.
For reference purposes, apart from the HMHD computations, we computed also a solution within classical MHD framework
(without the Hall corrections). A constant magnetic diffusivity $\eta=0.005$ was used in the HMHD and MHD simulations
with the isotropic pressure model including the total energy conservation equation.

\begin{figure}[!htbp]
\begin{center}
\includegraphics[scale=0.32]{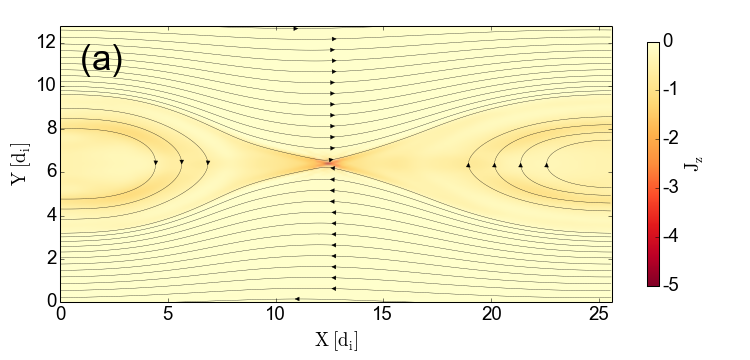}
\includegraphics[scale=0.32]{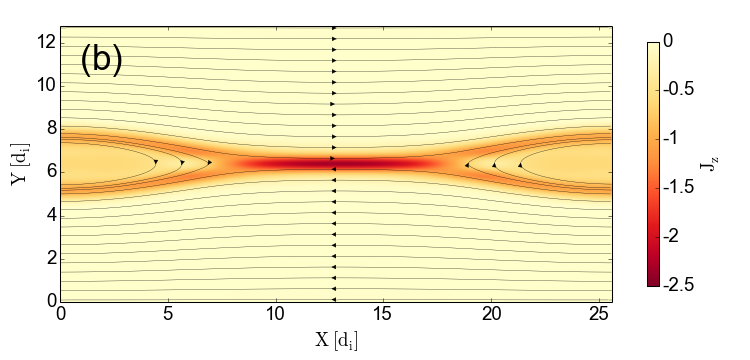}
\caption{Distributions of the out-of-plane component of the current density vector
for $t=30$ for (a) HMHD and (b) MHD models. Lines with arrows show the magnetic field lines.}
\label{f_rec_1}
\end{center}
\end{figure}

Fig. \ref{f_rec_1} shows the distributions of the out-of-plane component of the current density vector
for $t=30$ for the HMHD and MHD models. The distributions of the current density are
very similar to those presented in Ref. \cite{BirHes01}, in particular for the HMHD case [Fig. \ref{f_rec_1}(a)]
the current density is concentrated in the center of the simulation box.
Some papers (see, e.g. Ref. \cite{Totetal08}) reported an island in the center of the simulation box
in the HMHD case, this type of behavior appears in our simulations but for smaller values of the magnetic diffusivity (not shown here). Reconnection
along elongated current sheet for the MHD case [Fig. \ref{f_rec_1}(b)] is consistent with the Sweet-Parker model predictions.
At the left- and right-hand side of the simulation box in Fig. \ref{f_rec_1} one can see some gradients
that are related to periodic boundary conditions and can be seen in other simulations of
the magnetic reconnection sites (see e.g. \cite{Casetal05}). 

\begin{figure}[!htbp]
\begin{center}
\includegraphics[scale=0.35]{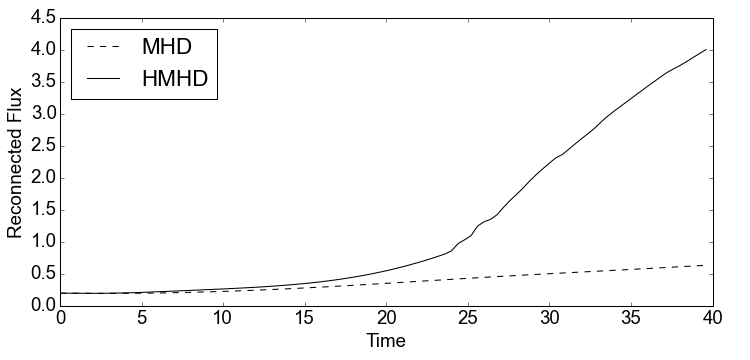}
\caption{Results of testing of the reconnected flux dependence on time for the HMHD and MHD simulations of
magnetic reconnection.}
\label{f_rec_2}
\end{center}
\end{figure}

The time evolution of the reconnected flux $\int_{L_x/2}^{L_x} B_y(x,y=L_y/2) dx$ is shown in Fig. \ref{f_rec_2}.
The HMHD flux is several times larger than the MHD flux, as typically obtained in this type of simulations
\cite{Biretal01,Hub03,Totetal08}. One should note that the time dependence of the reconnected
flux is quantitatively very similar to results presented in Ref. \cite{Biretal01} for both the HMHD and the MHD cases.

\subsection{Firehose instability}
Pressure anisotropy may provide free energy for the growth of instabilities
in plasmas. One of the examples is the firehose instability growing when
$\Lambda=\beta_\parallel-\beta_\bot-2>0$
\cite{Kunetal14,WanHau10,KenSag67,WanHau03}. We present results for 2D
simulation box, where the domain size is $L_x=L_y=64 \, d_i$ and for 3D simulations where
$L_x=L_y=L_z=64 \, d_i$. The mean magnetic field $\mathbf{B}/B_0=(1,0,0)$ is set up along the $x$ axis. Periodic boundary
conditions are applied in all directions. In the initial condition 
low-amplitude $\delta \mathbf{B}$ and $\delta \mathbf{u}$ fluctuations with randomized
phases are set up, the fluctuations are not correlated initially.
In tests presented in this subsection the magnetic diffusivity is $\eta=0$.

\begin{figure}[!htbp]
\begin{center}
\includegraphics[scale=0.32]{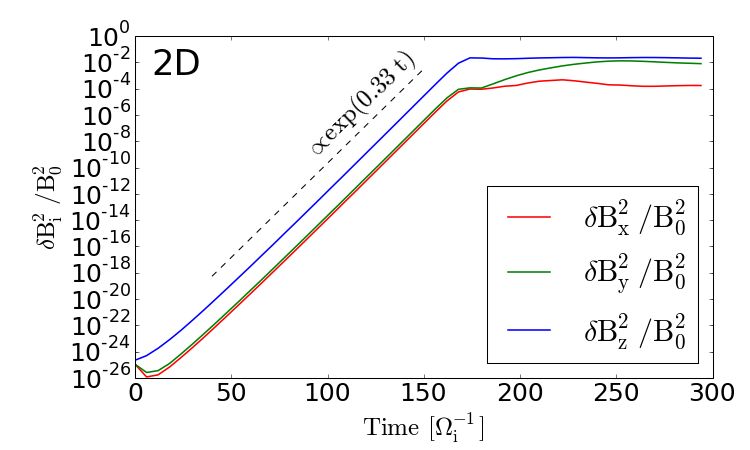}
\includegraphics[scale=0.32]{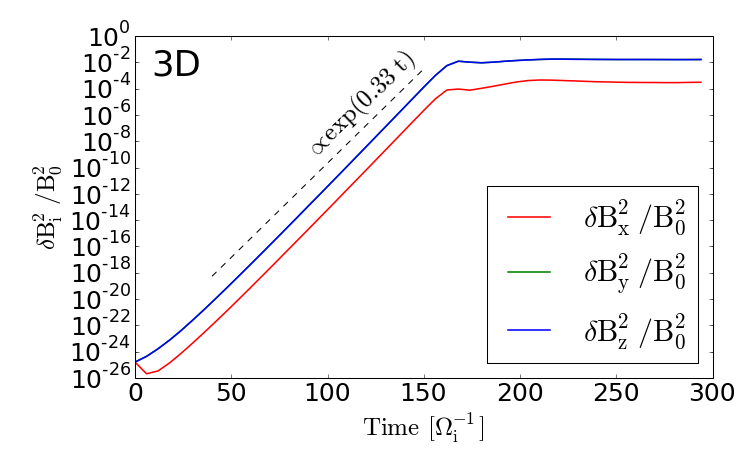}
\caption{Growth of the amplitude of the firehose instability fluctuations for
2D (on the left) and 3D (on the right) simulations for $\beta_\parallel=5$ and $\Lambda_0=0.2$.}
\label{f_fh_1}
\end{center}
\end{figure}

Fig. \ref{f_fh_1} shows the growth of the amplitude of fluctuations from the initial noise
for $\beta_\parallel=5$ and $\Lambda_0=0.2$ (in the initial condition) for 2D and 3D simulations.
The square of the amplitude $\delta B_z^2/B_0^2$ grows exponentially approx. 17 orders of magnitude until
$\delta B_z^2/B_0^2 \sim 0.01$ is reached, where $\delta B_z \gg \delta B_y \approx \delta B_x$.
One should note that the saturation
level is consistent with the predictions of the quasilinear theory \cite{Schetal08}
and similar to the saturation amplitude obtained in 2D high-$\beta$ kinetic simulation results
reported in Ref. \cite{Kunetal14b}.

\begin{figure}[!htbp]
\begin{center}
\includegraphics[scale=0.32]{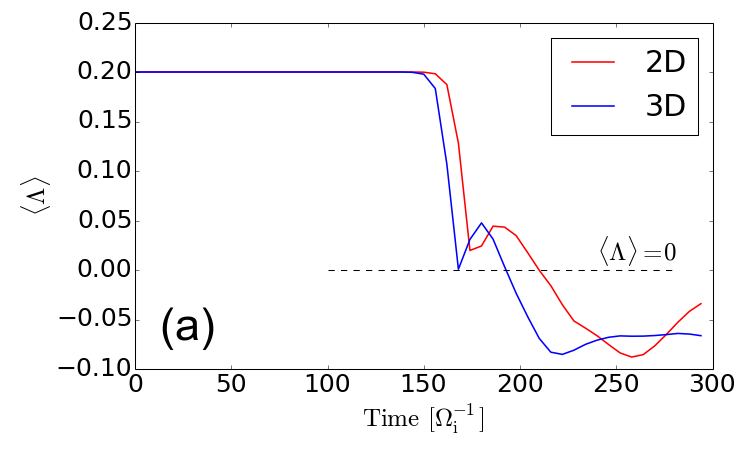}
\includegraphics[scale=0.32]{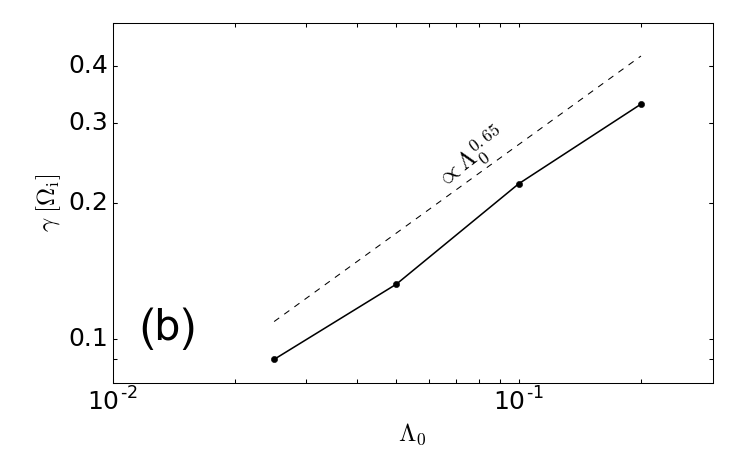}
\caption{Characteristics of the simulated firehose instability: (a)
dependence of the average anisotropy $\langle \Lambda \rangle$ on
time for 2D and 3D simulations and (b) dependence of the growth rate $\gamma$
on the initial anisotropy level $\Lambda_0$.}
\label{f_fh_2}
\end{center}
\end{figure}

Fig. \ref{f_fh_2}(a) shows that during the growth of the fluctuations the average pressure anisotropy
level $\langle \Lambda \rangle$ drops just below the
threshold $\Lambda=0$ for the firehose instability. Fig. \ref{f_fh_2}(b) shows results of testing the
dependence of the growth rate $\gamma$ on the initial pressure anisotropy $\Lambda_0$
for constant $\beta_\parallel=5$ in 2D simulations. A power-law relationship is obtained with $\gamma \propto \Lambda_0^{0.65}$.

\begin{figure}[!htbp]
\begin{center}
\includegraphics[scale=0.25]{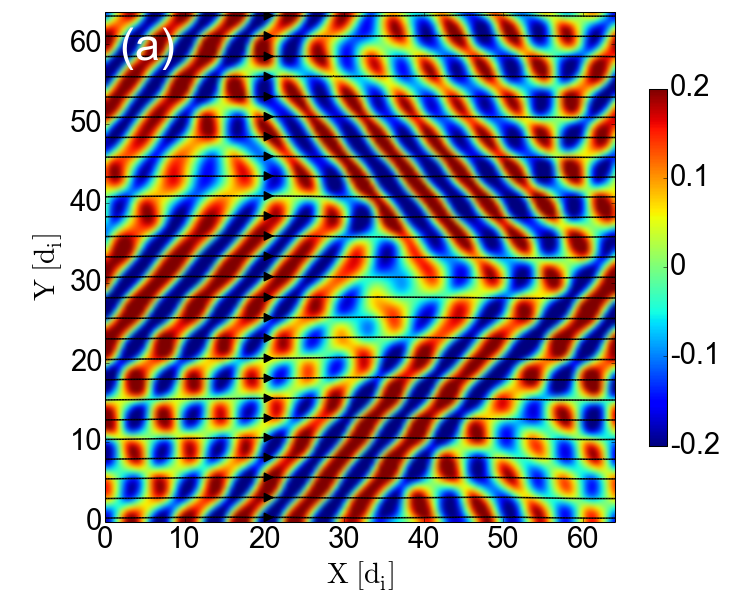}
\includegraphics[scale=0.25]{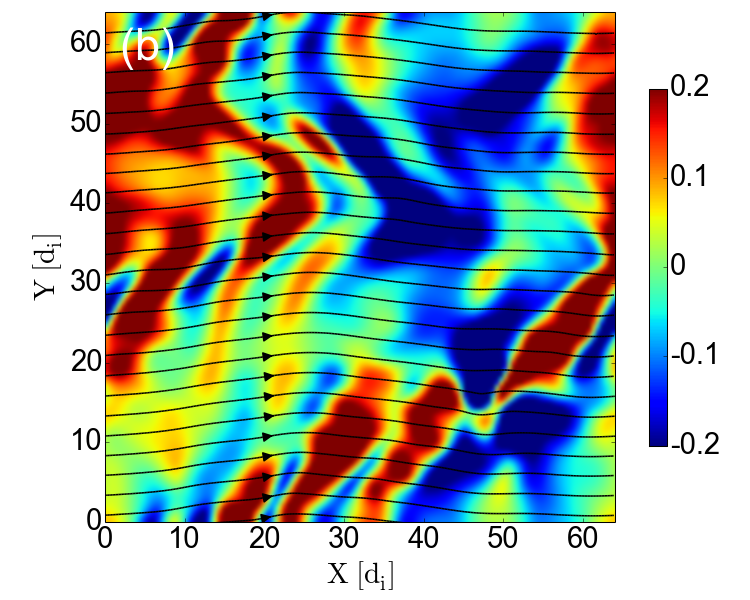}
\caption{Spatial distribution of the magnetic field component
$B_z$ for 2D simulations of the firehose instability for (a) $t=180 \, \Omega_i^{-1}$ and (b) $t=240 \, \Omega_i^{-1}$.
Black lines represent the magnetic field lines.}
\label{f_fh_3}
\end{center}
\end{figure}

\begin{figure}[!htbp]
\begin{center}
\includegraphics[scale=0.32]{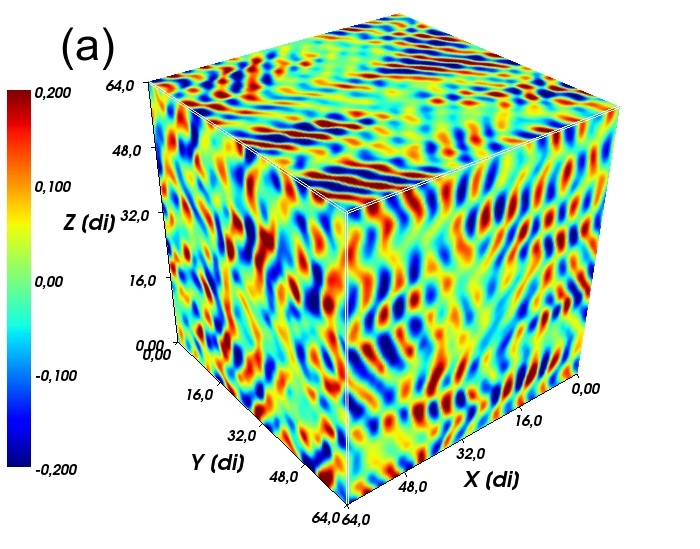}
\includegraphics[scale=0.32]{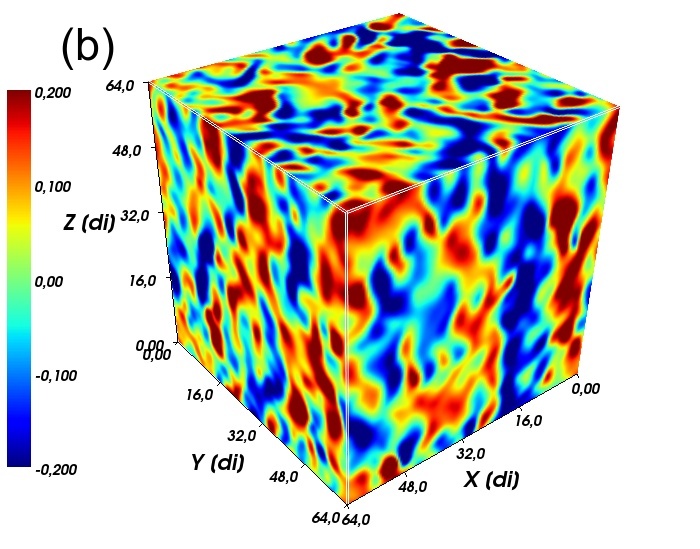}
\caption{Spatial distribution of the magnetic field component
$B_z$ for 3D simulations of the firehose instability for (a) $t=180 \, \Omega_i^{-1}$ and (b) $t=240 \, \Omega_i^{-1}$.}
\label{f_fh_4}
\end{center}
\end{figure}

\begin{figure}[!htbp]
\begin{center}
\includegraphics[scale=0.3]{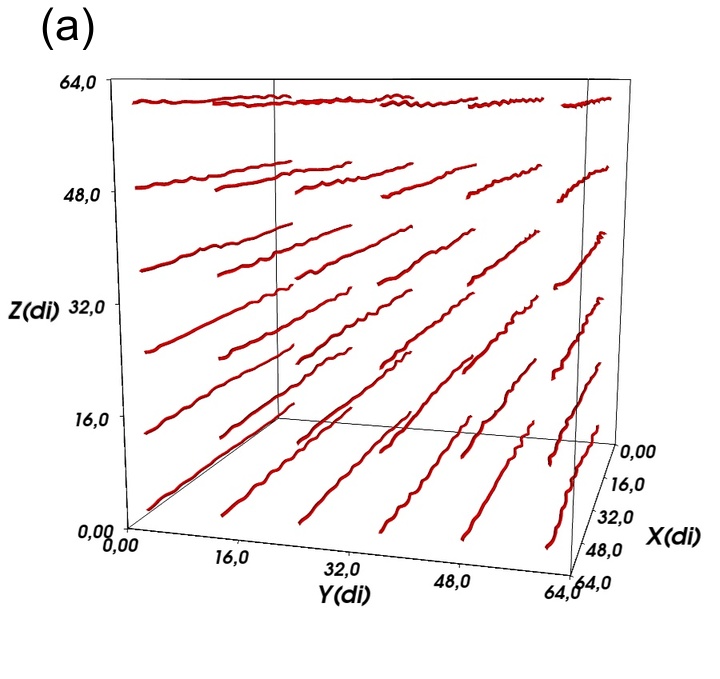}
\includegraphics[scale=0.3]{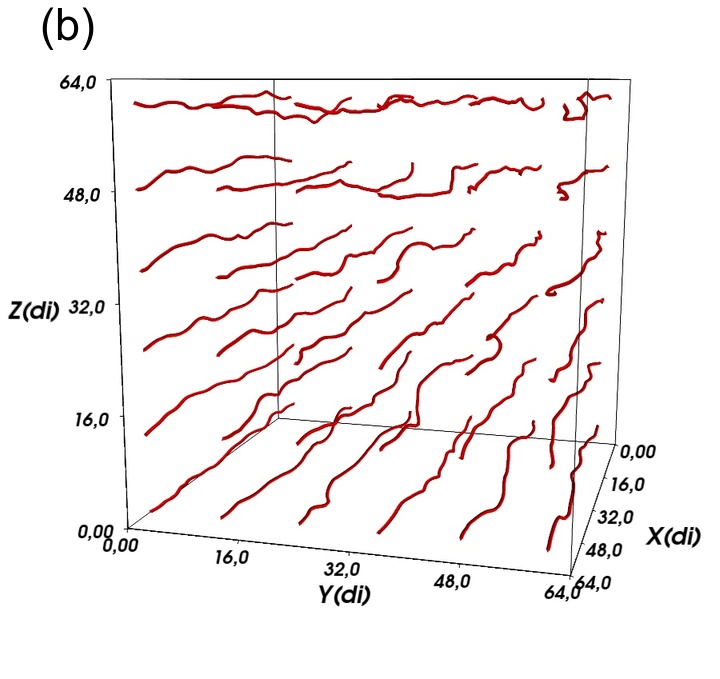}
\caption{Spatial structure of magnetic field lines for 3D simulations of the firehose instability for
(a) $t=180 \, \Omega_i^{-1}$ and (b) $t=240 \, \Omega_i^{-1}$.}
\label{f_fh_4_lines}
\end{center}
\end{figure}

As illustrated in Fig. \ref{f_fh_3} (for 2D) and Fig. \ref{f_fh_4} (for 3D) the
fastest growing mode is oblique with respect to the mean magnetic field
(oriented along the $x$ axis).
Initially locally regular distribution of $B_z$ (shown in Figs. \ref{f_fh_3}(a)
and \ref{f_fh_4}(a) for $t=180 \, \Omega_i^{-1}$) gradually changes in
time towards a more turbulent state (shown in Figs. \ref{f_fh_3}(b) and
\ref{f_fh_4}(b) for $t=240 \, \Omega_i^{-1}$).
The magnetic field lines can be seen to be weakly perturbed
in the simulation plane in 2D case, because fluctuations of the $B_z$
(out-of-plane) component grow preferentially in this case as seen in Fig. \ref{f_fh_1}.
The perturbation of the magnetic field lines in the 3D case is visualized
in Fig. \ref{f_fh_4_lines}. The behavior of the
firehose fluctuations is generally similar to results of 2D high-$\beta$
kinetic simulations reported in Ref. \cite{Kunetal14}. To our knowledge,
we present in our paper the first results of three-dimensional structure
of the firehose instability fluctuations.

One should note that fluid models give the same threshold for the parallel firehose
instability as the kinetic models, but a more stringent threshold is obtained
within kinetic theory for the oblique firehose instability \cite{Heletal06,HelMat00,HelMat01}.
This leads typically to preferential growth of the oblique mode, when the parallel and the oblique
instabilities compete. In our simulations within the HMHD framework, the oblique mode
grows in the system, which is similar to kinetic models and suggests that the
HMHD model may capture some elements of the oblique-parallel firehose mode competition.
Systematic analysis of the dispersion relation would be interesting in this context.
To our knowledge, only the parallel mode of the firehose instability in HMHD has been systematically
investigated \cite{WanHau10}.

\section{Summary}

We present a second-order accurate solver for the HMHD equations with
anisotropic or isotropic thermal pressure. Both explicit energy
conservation equation and polytropic state equations can be used in this
approach as the closure for the HMHD model. The implemented code was
validated using test problems previously described in the literature: the
magnetic reconnection process and the growth of the firehose instability.
Additionally, we propose a new validation method for the HMHD codes based
on solitary waves that provides a possibility of quantitative testing in
nonlinear regime as a complementary
approach to standard tests using small-amplitude whistler waves.
Quantitative tests of the accuracy and the performance of the implemented code
show the fidelity of the proposed approach. It is demonstrated that the
efficiency of the proposed approach and its implementation are 
sufficient for three-dimensional simulations within the HMHD framework.

The present formulation of the numerical scheme for the HMHD equations is
purely explicit and can be accelerated by using, e.g. a subcycling procedure for the
integration of the equation for the magnetic field transport as proposed
in Ref. \cite{Hub03}. Another possibility of acceleration of the
computations includes applying an implicit scheme, where stability of the code is
not constrained by the CFL condition for whistler waves at the expense of
the accuracy of the computations. In these contexts the proposed method
of validation based on solitary waves can be particularly
useful for testing the accuracy of the modified approaches. Since in some
aspects the HMHD model can be considered as a low-beta approximation for
kinetic models, the testing method based on solitary waves presented in
this paper has also conceivable applications for testing/validation of kinetic
(particle-in-cell or Vlasov-Maxwell) models.

\section*{Acknowledgments}
This work has been supported by the Polish National Science Centre
(DEC-2012/05/B/ST9/03916). One of the authors (M.S.) of this project has
received funding from the European Union's Horizon 2020 research and
innovation programme under the Marie Sklodowska-Curie grant agreement No
657251 (ASTROMULTISCALE). The discussion presented in the paper reflects
only the authors' view and the European Commission is not responsible for
any use that may be made of the information it contains. M.S. wishes to
acknowledge support from the International Space Science Institute for the
team ``Facing the Most Pressing Challenges to Our Understanding of the
Heliosheath and its Outer Boundaries''.

\section*{References}


\begin{thebibliography}{10}
\expandafter\ifx\csname url\endcsname\relax
  \def\url#1{\texttt{#1}}\fi
\expandafter\ifx\csname urlprefix\endcsname\relax\def\urlprefix{URL }\fi
\expandafter\ifx\csname href\endcsname\relax
  \def\href#1#2{#2} \def\path#1{#1}\fi

\bibitem{Hub95}
J.~D. {Huba}, {Hall magnetohydrodynamics in space and laboratory plasmas},
  Phys. Plasmas 2 (1995) 2504--2513.
\newblock \href {http://dx.doi.org/10.1063/1.871212}
  {\path{doi:10.1063/1.871212}}.

\bibitem{Biretal01}
J.~{Birn}, J.~F. {Drake}, M.~A. {Shay}, B.~N. {Rogers}, R.~E. {Denton},
  M.~{Hesse}, M.~{Kuznetsova}, Z.~W. {Ma}, A.~{Bhattacharjee}, A.~{Otto}, P.~L.
  {Pritchett}, {Geospace Environmental Modeling (GEM) magnetic reconnection
  challenge}, J.~Geophys. Res. 106 (2001) 3715--3720.
\newblock \href {http://dx.doi.org/10.1029/1999JA900449}
  {\path{doi:10.1029/1999JA900449}}.

\bibitem{MaBha01}
Z.~W. {Ma}, A.~{Bhattacharjee}, {Hall magnetohydrodynamic reconnection: The
  Geospace Environment Modeling challenge}, J.~Geophys. Res. 106 (2001)
  3773--3782.
\newblock \href {http://dx.doi.org/10.1029/1999JA001004}
  {\path{doi:10.1029/1999JA001004}}.

\bibitem{HubRud04}
J.~D. {Huba}, L.~I. {Rudakov}, {Hall Magnetic Reconnection Rate}, Phys. Rev.
  Lett. 93~(17) (2004) 175003.
\newblock \href {http://dx.doi.org/10.1103/PhysRevLett.93.175003}
  {\path{doi:10.1103/PhysRevLett.93.175003}}.

\bibitem{Sta04}
K.~{Stasiewicz}, {Reinterpretation of mirror modes as trains of slow
  magnetosonic solitons}, Geophys. Res. Lett. 31 (2004) 21804.
\newblock \href {http://dx.doi.org/10.1029/2004GL021282}
  {\path{doi:10.1029/2004GL021282}}.

\bibitem{Sta04b}
K.~{Stasiewicz}, {Theory and Observations of Slow-Mode Solitons in Space
  Plasmas}, Phys. Rev. Lett. 93~(12) (2004) 125004.
\newblock \href {http://dx.doi.org/10.1103/PhysRevLett.93.125004}
  {\path{doi:10.1103/PhysRevLett.93.125004}}.

\bibitem{Sta05}
K.~{Stasiewicz}, {Nonlinear Alfv{\'e}n, magnetosonic, sound, and electron
  inertial waves in fluid formalism}, J.~Geophys. Res. 110 (2005) 3220.
\newblock \href {http://dx.doi.org/10.1029/2004JA010852}
  {\path{doi:10.1029/2004JA010852}}.

\bibitem{Stretal11}
M.~{Strumik}, K.~{Stasiewicz}, C.~Z. {Cheng}, B.~{Thid{\'e}}, {Evolution of
  large-scale magnetosonic structures to trains of solitary waves}, J.~Geophys.
  Res. 116 (2011) 7209.
\newblock \href {http://dx.doi.org/10.1029/2011JA016565}
  {\path{doi:10.1029/2011JA016565}}.

\bibitem{Minetal05}
P.~D. {Mininni}, D.~O. {G{\'o}mez}, S.~M. {Mahajan}, {Direct Simulations of
  Helical Hall-MHD Turbulence and Dynamo Action}, Astrophys. J. 619 (2005)
  1019--1027.
\newblock \href {http://dx.doi.org/10.1086/426534} {\path{doi:10.1086/426534}}.

\bibitem{DmiMat06}
P.~{Dmitruk}, W.~H. {Matthaeus}, {Test particle acceleration in
  three-dimensional Hall MHD turbulence}, J.~Geophys. Res. 111~(10) (2006)
  12110.
\newblock \href {http://dx.doi.org/10.1029/2006JA011988}
  {\path{doi:10.1029/2006JA011988}}.

\bibitem{Kasetal02}
J.~C. {Kasper}, A.~J. {Lazarus}, S.~P. {Gary}, {Wind/SWE observations of
  firehose constraint on solar wind proton temperature anisotropy}, Geophys.
  Res. Lett. 29 (2002) 1839.
\newblock \href {http://dx.doi.org/10.1029/2002GL015128}
  {\path{doi:10.1029/2002GL015128}}.

\bibitem{Heletal06}
P.~{Hellinger}, P.~{Tr{\'a}vn{\'{\i}}{\v c}ek}, J.~C. {Kasper}, A.~J.
  {Lazarus}, {Solar wind proton temperature anisotropy: Linear theory and
  WIND/SWE observations}, Geophys. Res. Lett. 33 (2006) 9101.
\newblock \href {http://dx.doi.org/10.1029/2006GL025925}
  {\path{doi:10.1029/2006GL025925}}.

\bibitem{Matetal07}
L.~{Matteini}, S.~{Landi}, P.~{Hellinger}, F.~{Pantellini}, M.~{Maksimovic},
  M.~{Velli}, B.~E. {Goldstein}, E.~{Marsch}, {Evolution of the solar wind
  proton temperature anisotropy from 0.3 to 2.5 AU}, Geophys. Res. Lett. 34
  (2007) 20105.
\newblock \href {http://dx.doi.org/10.1029/2007GL030920}
  {\path{doi:10.1029/2007GL030920}}.

\bibitem{Baletal09}
S.~D. {Bale}, J.~C. {Kasper}, G.~G. {Howes}, E.~{Quataert}, C.~{Salem},
  D.~{Sundkvist}, {Magnetic Fluctuation Power Near Proton Temperature
  Anisotropy Instability Thresholds in the Solar Wind}, Phys. Rev. Lett.
  103~(21) (2009) 211101.
\newblock \href {http://dx.doi.org/10.1103/PhysRevLett.103.211101}
  {\path{doi:10.1103/PhysRevLett.103.211101}}.

\bibitem{Schetal05}
A.~A. {Schekochihin}, S.~C. {Cowley}, R.~M. {Kulsrud}, G.~W. {Hammett},
  P.~{Sharma}, {Plasma Instabilities and Magnetic Field Growth in Clusters of
  Galaxies}, Astrophys. J. 629 (2005) 139--142.
\newblock \href {http://dx.doi.org/10.1086/431202} {\path{doi:10.1086/431202}}.

\bibitem{HelTra08}
P.~{Hellinger}, P.~M. {Tr{\'a}vn{\'{\i}}{\v c}ek}, {Oblique proton fire hose
  instability in the expanding solar wind: Hybrid simulations}, J.~Geophys.
  Res. 113~(A12) (2008) 10109.
\newblock \href {http://dx.doi.org/10.1029/2008JA013416}
  {\path{doi:10.1029/2008JA013416}}.

\bibitem{Schetal08}
A.~A. {Schekochihin}, S.~C. {Cowley}, R.~M. {Kulsrud}, M.~S. {Rosin},
  T.~{Heinemann}, {Nonlinear Growth of Firehose and Mirror Fluctuations in
  Astrophysical Plasmas}, Phys. Rev. Lett. 100~(8) (2008) 081301.
\newblock \href {http://dx.doi.org/10.1103/PhysRevLett.100.081301}
  {\path{doi:10.1103/PhysRevLett.100.081301}}.

\bibitem{Kunetal14}
M.~W. {Kunz}, A.~A. {Schekochihin}, J.~M. {Stone}, {Firehose and Mirror
  Instabilities in a Collisionless Shearing Plasma}, Phys. Rev. Lett. 112~(20)
  (2014) 205003.
\newblock \href {http://dx.doi.org/10.1103/PhysRevLett.112.205003}
  {\path{doi:10.1103/PhysRevLett.112.205003}}.

\bibitem{Seretal14}
S.~{Servidio}, K.~T. {Osman}, F.~{Valentini}, D.~{Perrone}, F.~{Califano},
  S.~{Chapman}, W.~H. {Matthaeus}, P.~{Veltri}, {Proton Kinetic Effects in
  Vlasov and Solar Wind Turbulence}, Astrophys. J. Lett. 781 (2014) L27.
\newblock \href {http://dx.doi.org/10.1088/2041-8205/781/2/L27}
  {\path{doi:10.1088/2041-8205/781/2/L27}}.

\bibitem{Hub03}
J.~D. {Huba}, {Hall Magnetohydrodynamics - A Tutorial}, in: J.~{B{\"u}chner},
  C.~{Dum}, M.~{Scholer} (Eds.), Space Plasma Simulation, Vol. 615 of Lecture
  Notes in Physics, Berlin Springer Verlag, 2003, pp. 166--192.

\bibitem{ChaKno03}
L.~{Chac{\'o}n}, D.~A. {Knoll}, {A 2D high-/{$\beta$} Hall MHD implicit
  nonlinear solver}, J.~Comp. Phys. 188 (2003) 573--592.
\newblock \href {http://dx.doi.org/10.1016/S0021-9991(03)00193-1}
  {\path{doi:10.1016/S0021-9991(03)00193-1}}.

\bibitem{Lavetal09}
D.~{Laveder}, D.~{Borgogno}, T.~{Passot}, P.~L. {Sulem}, {On a semi-implicit
  scheme for spectral simulations of dispersive magnetohydrodynamics}, Computer
  Physics Communications 180 (2009) 1860--1869.
\newblock \href {http://dx.doi.org/10.1016/j.cpc.2009.05.018}
  {\path{doi:10.1016/j.cpc.2009.05.018}}.

\bibitem{Totetal08}
G.~{T{\'o}th}, Y.~{Ma}, T.~I. {Gombosi}, {Hall magnetohydrodynamics on
  block-adaptive grids}, Journal of Computational Physics 227 (2008)
  6967--6984.
\newblock \href {http://dx.doi.org/10.1016/j.jcp.2008.04.010}
  {\path{doi:10.1016/j.jcp.2008.04.010}}.

\bibitem{KraTri73}
N.~A. {Krall}, A.~W. {Trivelpiece}, {Principles of plasma physics},
  McGraw-Hill, 1973.

\bibitem{Leetal09}
A.~{Le}, J.~{Egedal}, W.~{Daughton}, W.~{Fox}, N.~{Katz}, {Equations of State
  for Collisionless Guide-Field Reconnection}, Phys. Rev. Lett. 102~(8) (2009)
  085001.
\newblock \href {http://dx.doi.org/10.1103/PhysRevLett.102.085001}
  {\path{doi:10.1103/PhysRevLett.102.085001}}.

\bibitem{Egeetal13}
J.~{Egedal}, A.~{Le}, W.~{Daughton}, {A review of pressure anisotropy caused by
  electron trapping in collisionless plasma, and its implications for magnetic
  reconnection}, Phys. Plasmas 20~(6) (2013) 061201.
\newblock \href {http://dx.doi.org/10.1063/1.4811092}
  {\path{doi:10.1063/1.4811092}}.

\bibitem{Bit04}
J.~A. {Bittencourt}, {Fundamentals of Plasma Physics}, Springer-Verlag, 2004.

\bibitem{Hauetal93}
L.-N. {Hau}, T.-D. {Phan}, B.~U.~O. {Sonnerup}, G.~{Paschmann},
  {Double-polytropic closure in the magentosheath}, Geophys. Res. Lett. 20
  (1993) 2255--2258.
\newblock \href {http://dx.doi.org/10.1029/93GL02491}
  {\path{doi:10.1029/93GL02491}}.

\bibitem{Cheetal57}
G.~F. {Chew}, M.~L. {Goldberger}, F.~E. {Low}, {The Boltzmann Equation and the
  One-Fluid Hydromagnetic Equations in the Absence of Particle Collisions},
  Royal Society of London Proceedings Series A 236 (1956) 112--118.
\newblock \href {http://dx.doi.org/10.1098/rspa.1956.0116}
  {\path{doi:10.1098/rspa.1956.0116}}.

\bibitem{Haketal06}
A.~{Hakim}, J.~{Loverich}, U.~{Shumlak}, {A high resolution wave propagation
  scheme for ideal Two-Fluid plasma equations}, J.~Comp. Phys. 219 (2006)
  418--442.
\newblock \href {http://dx.doi.org/10.1016/j.jcp.2006.03.036}
  {\path{doi:10.1016/j.jcp.2006.03.036}}.

\bibitem{Baletal16}
D.~S. {Balsara}, T.~{Amano}, S.~{Garain}, J.~{Kim}, {A high-order relativistic
  two-fluid electrodynamic scheme with consistent reconstruction of
  electromagnetic fields and a multidimensional Riemann solver for
  electromagnetism}, J.~Comp. Phys. 318 (2016) 169--200.
\newblock \href {http://dx.doi.org/10.1016/j.jcp.2016.05.006}
  {\path{doi:10.1016/j.jcp.2016.05.006}}.

\bibitem{Ama16}
T.~Amano, A second-order divergence-constrained multidimensional numerical
  scheme for relativistic two-fluid electrodynamics, Astrophys. J.In press
  "{arXiv:1607.08487}".

\bibitem{Wanetal15}
L.~{Wang}, A.~H. {Hakim}, A.~{Bhattacharjee}, K.~{Germaschewski}, {Comparison
  of multi-fluid moment models with particle-in-cell simulations of
  collisionless magnetic reconnection}, Phys. Plasmas 22~(1) (2015) 012108.
\newblock \href {http://dx.doi.org/10.1063/1.4906063}
  {\path{doi:10.1063/1.4906063}}.

\bibitem{WanHau10}
B.-J. {Wang}, L.-N. {Hau}, {Parallel proton fire hose instability in gyrotropic
  Hall MHD model}, J.~Geophys. Res. 115 (2010) 4105.
\newblock \href {http://dx.doi.org/10.1029/2009JA014947}
  {\path{doi:10.1029/2009JA014947}}.

\bibitem{McKetal04}
J.~F. {McKenzie}, E.~{Dubinin}, K.~{Sauer}, T.~B. {Doyle}, {The application of
  the constants of motion to nonlinear stationary waves in complex plasmas: a
  unified fluid dynamic viewpoint}, J.~Plasma Phys. 70 (2004) 431--462.
\newblock \href {http://dx.doi.org/10.1017/S0022377803002654}
  {\path{doi:10.1017/S0022377803002654}}.

\bibitem{BalSpi99}
D.~S. {Balsara}, D.~S. {Spicer}, {A Staggered Mesh Algorithm Using High Order
  Godunov Fluxes to Ensure Solenoidal Magnetic Fields in Magnetohydrodynamic
  Simulations}, J.~Comp. Phys. 149 (1999) 270--292.
\newblock \href {http://dx.doi.org/10.1006/jcph.1998.6153}
  {\path{doi:10.1006/jcph.1998.6153}}.

\bibitem{KurTad00}
A.~{Kurganov}, E.~{Tadmor}, {New High-Resolution Central Schemes for Nonlinear
  Conservation Laws and Convection-Diffusion Equations}, J.~Comp. Phys. 160
  (2000) 241--282.
\newblock \href {http://dx.doi.org/10.1006/jcph.2000.6459}
  {\path{doi:10.1006/jcph.2000.6459}}.

\bibitem{Tot00}
G.~{T{\'o}th}, {The \mbox{${\nabla}{\cdot}B=0$} Constraint in Shock-Capturing
  Magnetohydrodynamics Codes}, J.~Comp. Phys. 161 (2000) 605--652.
\newblock \href {http://dx.doi.org/10.1006/jcph.2000.6519}
  {\path{doi:10.1006/jcph.2000.6519}}.

\bibitem{Yee66}
K.~{Yee}, {Numerical solution of inital boundary value problems involving
  maxwell's equations in isotropic media}, IEEE Transactions on Antennas and
  Propagation 14 (1966) 302--307.
\newblock \href {http://dx.doi.org/10.1109/TAP.1966.1138693}
  {\path{doi:10.1109/TAP.1966.1138693}}.

\bibitem{EvaHaw88}
C.~R. {Evans}, J.~F. {Hawley}, {Simulation of magnetohydrodynamic flows - A
  constrained transport method}, Astrophys. J. 332 (1988) 659--677.
\newblock \href {http://dx.doi.org/10.1086/166684} {\path{doi:10.1086/166684}}.

\bibitem{Preetal92}
W.~H. {Press}, S.~A. {Teukolsky}, W.~T. {Vetterling}, B.~P. {Flannery},
  {Numerical recipes in C. The art of scientific computing}, Cambridge:
  University Press, 1992, 1992.

\bibitem{RamVar11}
P.~Ramachandran, G.~Varoquaux, Mayavi: 3d visualization of scientific data,
  Computing in Science \& Engineering 13~(2) (2011) 40--51.
\newblock \href {http://dx.doi.org/10.1109/MCSE.2011.35}
  {\path{doi:10.1109/MCSE.2011.35}}.

\bibitem{BirHes01}
J.~{Birn}, M.~{Hesse}, {Geospace Environment Modeling (GEM) magnetic
  reconnection challenge: Resistive tearing, anisotropic pressure and hall
  effects}, J.~Geophys. Res. 106 (2001) 3737--3750.
\newblock \href {http://dx.doi.org/10.1029/1999JA001001}
  {\path{doi:10.1029/1999JA001001}}.

\bibitem{Casetal05}
P.~A. {Cassak}, M.~A. {Shay}, J.~F. {Drake}, {Catastrophe Model for Fast
  Magnetic Reconnection Onset}, Phys. Rev. Lett. 95~(23) (2005) 235002.
\newblock \href {http://dx.doi.org/10.1103/PhysRevLett.95.235002}
  {\path{doi:10.1103/PhysRevLett.95.235002}}.

\bibitem{KenSag67}
C.~F. {Kennel}, R.~Z. {Sagdeev}, {Collisionless shock waves in high {$\beta$}
  plasmas: 1}, J.~Geophys. Res. 72 (1967) 3303--3326.
\newblock \href {http://dx.doi.org/10.1029/JZ072i013p03303}
  {\path{doi:10.1029/JZ072i013p03303}}.

\bibitem{WanHau03}
B.~J. {Wang}, L.~N. {Hau}, {MHD aspects of fire-hose type instabilities},
  J.~Geophys. Res. 108 (2003) 1463.
\newblock \href {http://dx.doi.org/10.1029/2003JA009986}
  {\path{doi:10.1029/2003JA009986}}.

\bibitem{Kunetal14b}
M.~W. {Kunz}, J.~M. {Stone}, X.-N. {Bai}, {Pegasus: A new hybrid-kinetic
  particle-in-cell code for astrophysical plasma dynamics}, J.~Comp. Phys. 259
  (2014) 154--174.
\newblock \href {http://dx.doi.org/10.1016/j.jcp.2013.11.035}
  {\path{doi:10.1016/j.jcp.2013.11.035}}.

\bibitem{HelMat00}
P.~{Hellinger}, H.~{Matsumoto}, {New kinetic instability: Oblique Alfv{\'e}n
  fire hose}, J.~Geophys. Res. 105 (2000) 10519--10526.
\newblock \href {http://dx.doi.org/10.1029/1999JA000297}
  {\path{doi:10.1029/1999JA000297}}.

\bibitem{HelMat01}
P.~{Hellinger}, H.~{Matsumoto}, {Nonlinear competition between the whistler and
  Alfv{\'e}n fire hoses}, J.~Geophys. Res. 106 (2001) 13215--13218.
\newblock \href {http://dx.doi.org/10.1029/2001JA900026}
  {\path{doi:10.1029/2001JA900026}}.

\end{thebibliography}

\end{document}